\documentclass[11pt]{article}
\usepackage{fullpage,amssymb,amsmath,amsthm,url,graphicx}
\usepackage{orcidlink}
 
\graphicspath{{./},{Fig/}}

\newtheorem{Remark}{Remark}
\newtheorem{Theorem}{Theorem}
\newtheorem{Example}{Example}
\newtheorem{Definition}{Definition}
\newtheorem{Property}{Property}

\def\dx{\mathrm{d}x}
\def\Bhat{\mathrm{Bhat}}
\def\calP{\mathcal{P}}
\def\calQ{\mathcal{Q}}
\def\calX{\mathcal{X}}
\def\st{\ :\ }
\def\dt{\mathrm{d}t}
\def\dP{\mathrm{d}P}
\def\dQ{\mathrm{d}Q}
\def\dmu{\mathrm{d}\mu}
\def\bbR{\mathbb{R}}
\def\KL{\mathrm{KL}}

\def\bbN{\mathbb{N}}
\def\dnu{\mathrm{d}\nu}
\def\calE{\mathcal{E}}
\def\IS{\mathrm{IS}}
\def\erf{\mathrm{erf}}
\def\inner#1#2{{\langle #1,#2\rangle}}

\title{The duo Bregman and Fenchel-Young divergences\footnote{A revised and extended paper is published in the journal: ``Statistical divergences between densities of truncated exponential families with nested supports: Duo Bregman and duo Jensen divergences,'' Entropy 2022, 24(3), 421; \protect\url{https://doi.org/10.3390/e24030421}~\cite{nielsen2022statistical}.}}

\author{Frank Nielsen~\orcidlink{0000-0001-5728-0726}\\ Sony Computer Science Laboratories Inc.\\ \ \\ Tokyo, Japan}

\date{}

\begin{document}

\maketitle

\begin{abstract}
By calculating the Kullback-Leibler divergence between two probability measures belonging to different exponential families, we end up with a formula that generalizes the ordinary Fenchel-Young divergence.
Inspired by this formula, we define the duo Fenchel-Young divergence and report a majorization condition on its pair of generators which guarantees  that this divergence is always non-negative. 
The duo Fenchel-Young divergence is also equivalent to a duo Bregman divergence.
We show the use of these duo divergences by calculating the Kullback-Leibler divergence between densities of nested exponential families, and report a formula for the Kullback-Leibler divergence between truncated normal distributions.
Finally, we prove that the skewed Bhattacharyya distance between nested exponential families amounts to an equivalent skewed duo Jensen divergence.
\end{abstract}

{\noindent Keywords:} exponential family; statistical divergence; truncated normal distributions; centroids.

%%%%
\section{Introduction}
%%%%

%%%
\subsection{Exponential families}
%%%
Let $(\calX,\Sigma)$ be a measurable space, and consider a regular minimal  exponential family~\cite{EF-2019} $\calE$ of probability measures $P_\theta$ all dominated by a base measure $\mu$ ($P_\theta\ll \mu$):
$$
\calE=\{P_\theta\st \theta\in\Theta\}.
$$

The Radon-Nikodym derivatives of the probability measures $P_\theta$ with respect to $\mu$ can be written canonically as
$$
p_\theta(x)=\frac{\dP_\theta}{\dmu}(x)=\exp\left(\theta^\top t(x)-F(\theta)+k(x)\right),
$$
where $\theta$ denotes the natural parameter,  $t(x)$ the sufficient statistic, and $F(\theta)$ the log-normalizer~\cite{EF-2019} (or cumulant function).
The optional auxiliary term $k(x)$ allows to change the base measure $\mu$ into the measure $\nu$ such that
 $\frac{\dnu}{\dmu}(x)=e^{k(x)}$.
The distributions $P_\theta$ of the exponential family $\calE$ can be interpreted as distributions obtained by tilting the base measure $\mu$~\cite{efron2021computer}. Thus when $t(x)=x$, these natural exponential families~\cite{EF-2019} are also called tilted exponential families~\cite{hiejima1997interpretation} in the literature.

%%%%%
\subsection{Kullback-Leibler divergence between exponential family distributions}
%%%%
For two $\sigma$-finite probability measures $P$ and $Q$ on $(\calX,\Sigma)$ such that $P$ is dominated by $Q$ ($P \ll Q$), the Kullback-Leibler divergence between $P$ and $Q$ is defined by
$$
D_\KL[P:Q]=\int_\calX \log\frac{\dP}{\dQ}\, \dP=E_P\left[\log \frac{\dP}{\dQ}\right].
$$

When $P \not\ll Q$, we set $D_\KL[P:Q]=+\infty$.
Gibbs' inequality~\cite{cover1999elements} $D_\KL[P:Q]\geq 0$ shows that the KLD is always non-negative.  
The proof of Gibbs' inequality relies on Jensen's inequality and holds for the wide class of $f$-divergences~\cite{csiszar1964informationstheoretische} induced by convex generators $f(u)$:
$$
I_f[P:Q]=\int_\calX f\left(\frac{\dQ}{\dP}\right)\, \dP\geq f\left( \int_\calX \frac{\dQ}{\dP}\,\dP \right)\geq f(1). 
$$
The KLD is a $f$-divergence for the convex generator $f(u)=-\log u$.

%%%
\subsection{Kullback-Leibler divergence between exponential family densities}
%%%
It is well-known that the KLD between two distributions $P_{\theta_1}$ and $P_{\theta_2}$ of $\calE$ amounts to compute an equivalent Fenchel-Young divergence~\cite{azoury2001relative}:
$$
D_\KL[P_{\theta_1}:P_{\theta_2}]=\int_\calX p_{\theta_1}\log\frac{p_{\theta_1}}{p_{\theta_2}}\, \dmu(x)=Y_{F,F^*}(\theta_2,\eta_1),
$$
where $\eta=\nabla F(\theta)=E_{P_\theta}[t(x)]$ is the moment parameter~\cite{EF-2019}, and the Fenchel-Young divergence is defined for a pair of strictly convex conjugate functions~\cite{rockafellar2015convex} $F(\theta)$ and $F^*(\eta)$ related by the Legendre-Fenchel transform by 
$$
Y_{F,F^*}(\theta_1,\eta_2):=F(\theta_1)+F^*(\eta_2)-\theta_1^\top\eta_2.
$$

Amari (1985) first introduced this formula as the canonical divergence of dually flat spaces in information geometry~\cite{amari1985differential} (Eq. 3.21), and proved that the Fenchel-Young divergence is obtained as the KLD between densities belonging to  a same exponential family~\cite{amari1985differential} (Theorem 3.7).
Azoury and Warmuth expressed the KLD $D_\KL[P_{\theta_1}:P_{\theta_2}]$ using dual Bregman divergences in~\cite{azoury2001relative} (2001):
$$
D_\KL[P_{\theta_1}:P_{\theta_2}]=B_F(\theta_2:\theta_1)=B_{F^*}(\eta_1:\eta_2),
$$
where a Bregman divergence~\cite{bregman1967relaxation} $B_F(\theta_1:\theta_2)$ is defined for a strictly convex and differentiable generator $F(\theta)$ by:
$$
B_F(\theta_1:\theta_2):=F(\theta_1)-F(\theta_2)-(\theta_1-\theta_2)^\top \nabla F(\theta_2).
$$

\begin{figure}
\centering
\includegraphics[width=0.75\textwidth]{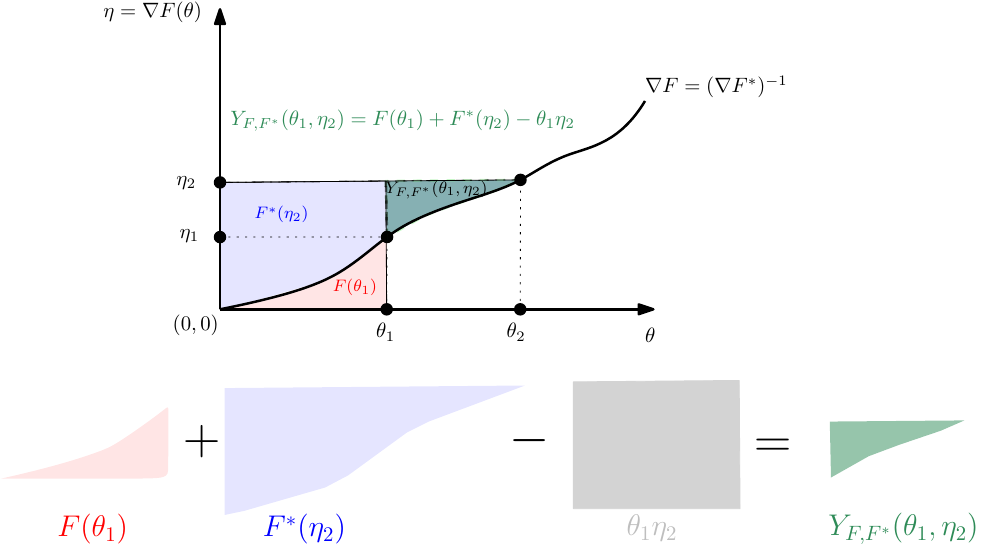}
\caption{Visualizing the Fenchel--Young divergence.}\label{fig:FY}
\end{figure}

Acharyya termed the divergence $Y_{F,F^*}$ the {Fenchel-Young divergence} in his PhD thesis~\cite{acharyya2013learning} (2013), and
 Blondel et al. called them {Fenchel-Young losses} (2020) in the context of machine learning~\cite{blondel2020learning} (Eq. 9 in Definition 2). It was also called by the author the Legendre-Fenchel divergence in~\cite{nielsen2020elementary}.
The Fenchel-Young divergence stems from the Fenchel-Young inequality~\cite{rockafellar2015convex}:
$$
F(\theta_1)+F^*(\eta_2)\geq \theta_1^\top\eta_2,
$$
with equality iff. $\eta_2=\nabla F(\theta_1)$.
Figure~\ref{fig:FY} visualizes the 1D Fenchel--Young divergence and gives a geometric proof that $Y_{F,F^*}(\theta_1,\eta_2)\geq 0$ with equality if and only if $\eta_2=F'(\theta_1)$.

\begin{figure}
\centering
\includegraphics[width=0.75\textwidth]{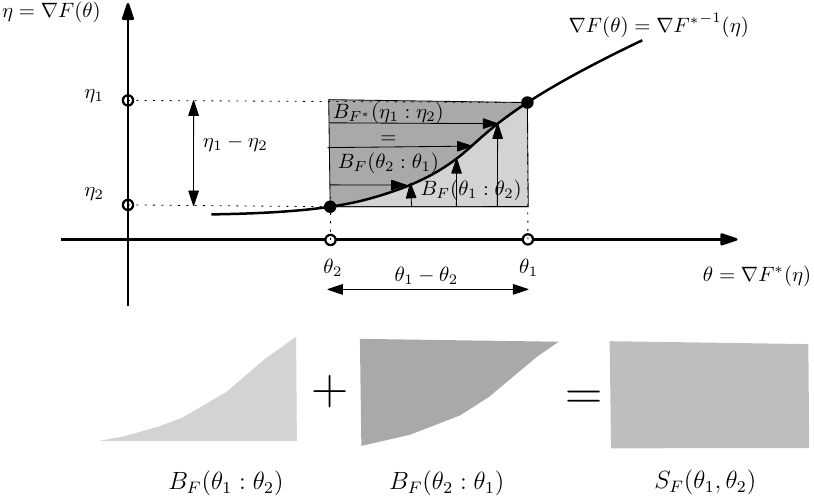}
\caption{Visualizing the sided and symmetrized Bregman divergences.}\label{fig:SBD}
\end{figure}

The symmetrized Kullback-Leibler divergence  $D_J[P_{\theta_1}:P_{\theta_2}]$ between two distributions $P_{\theta_1}$ and $P_{\theta_2}$ of $\calE$ is called the Jeffreys' divergence~\cite{jeffreys1998theory} and amounts to a symmetrized Bregman divergence~\cite{nielsen2009sided}:
\begin{eqnarray}
D_J[P_{\theta_1}:P_{\theta_2}] &=& D_\KL[P_{\theta_1}:P_{\theta_2}]+D_\KL[P_{\theta_2}:P_{\theta_1}],\\
&=& B_F(\theta_2:\theta_1)+B_F(\theta_1:\theta_2),\\
&=& (\theta_2-\theta_1)^\top (\eta_2-\eta_1):= S_F(\theta_1,\theta_2).
\end{eqnarray}
Notice that the Bregman divergence $B_F(\theta_1:\theta_2)$ can also be interpreted as a surface area:
\begin{equation}
B_F(\theta_1:\theta_2) = \int_{\theta_2}^{\theta_1} (F'(\theta)-F'(\theta_2)) \mathrm{d}\theta.
\end{equation}
Figure~\ref{fig:SBD} illustrates the sided and symmetrized Bregman divergences.

%%%
\subsection{Contributions and paper outline}
%%%

We recall in \S\ref{sec:kldefs} the formula obtained for the Kullback-Leibler divergence between two exponential family densities equivalent to each other~\cite{JS-2021} (Eq.~\ref{eq:KLDEFs}).
Inspired by this formula, we give a definition of the duo Fenchel-Young divergence induced by a pair of strictly convex functions $F_1$ and $F_2$ (Definition~\ref{def:genyf}) in \S\ref{sec:duoyf}, and proves that the divergence is always non-negative provided that $F_1$ upper bounds $F_2$.
We then define the duo Bregman divergence  (Definition~\ref{def:genbd}) corresponding to the duo Fenchel-Young divergence.
In \S\ref{sec:kldtrunc}, we show that the Kullback-Leibler divergence between a truncated density and a density of a same parametric exponential family amounts to a duo Fenchel-Young divergence or equivalently to a Bregman divergence  on swapped parameters (Theorem~\ref{thm:KLDnestedEF}). As an example, we report a formula for the Kullback-Leibler divergence between truncated normal distributions (Example~\ref{ex:truncnormal}).
In \S\ref{sec:bhat}, we further consider the skewed Bhattacharyya distance between nested exponential family densities and prove that it amounts to a duo Jensen divergence (Theorem~\ref{thm:BhatNEF}).
Finally, we conclude in~\S\ref{sec:concl}.

%%%
\section{Kullback-Leibler divergence between different exponential families}\label{sec:kldefs}
%%%

Consider now two exponential families~\cite{EF-2019} $\calP$ and $\calQ$ 
defined by their Radon-Nikodym derivatives with respect to two positive measures $\mu_{\calP}$ and $\mu_{\calQ}$ on  $(\calX,\Sigma)$:
\begin{eqnarray*}
\calP &=&\left\{ P_\theta\st \theta\in\Theta\right\},\\
\calQ &=&\left\{ Q_{\theta'}\st \theta'\in\Theta'\right\}.
\end{eqnarray*}
The corresponding natural parameter spaces are
\begin{eqnarray*}
\Theta &=&\left\{ \theta\in\bbR^{D} \st \int_\calX \exp(\theta^\top t_\calP(x)+k_\calP(x))\, \dmu_\calP(x)<\infty  \right\},\\
\Theta' &=&\left\{  \theta'\in\bbR^{D'} \st \int_\calX \exp({\theta'}^\top t_\calQ(x)+k_\calQ(x))\, \dmu_\calQ(x)<\infty  \right\},
\end{eqnarray*}
The order of $\calP$ is $D$,   $t_\calP(x)$ denotes the sufficient statistics of $P_\theta$, and $k_\calP(x)$ is a term to adjust/tilt the base measure $\mu_\calP$. 
Similarly, the order of $\calQ$ is $D'$,   $t_\calQ(x)$ denotes the sufficient statistics of $Q_{\theta'}$, and $k_\calQ(x)$ is an optional term to adjust the base measure $\mu_\calQ$.
Let $p_\theta$ and $q_{\theta'}$ denote the Radon-Nikodym with respect to the measure $\mu_\calP$ and $\mu_\calQ$, respectively:
\begin{eqnarray*}
p_\theta&=&\frac{\dP_\theta}{\dmu_\calP}  = \exp(\theta^\top t_\calP(x)-F_\calP(\theta)+k_\calP(x)),\\
q_{\theta'}&=&\frac{\dQ_{\theta'}}{\dmu_\calQ}  = \exp({\theta'}^\top t_\calQ(x)-F_\calQ(\theta')+k_\calQ(x)), 
\end{eqnarray*}
where $F_\calP(\theta)$ and $F_\calQ(\theta')$ denote the corresponding log-normalizers of $\calP$ and $\calQ$, respectively.
\begin{eqnarray*}
F_\calP(\theta)&=&\log \left(\int \exp(\theta^\top t_\calP(x)+k_\calP(x))\, \dmu_\calP(x)\right),\\
F_\calQ(\theta)&=&\log \left(\int \exp(\theta^\top t_\calQ(x)+k_\calQ(x))\, \dmu_\calQ(x)\right).
\end{eqnarray*}

The functions $F_\calP$ and $F_\calQ$ are strictly convex and real analytic~\cite{EF-2019}.
Hence, those functions are infinitely many times differentiable on their open natural parameter spaces.

Consider the KLD between $P_\theta\in\calP$ and $Q_{\theta'}\in\calQ$ such that $\mu_\calP=\mu_\calQ$ (and hence $P_\theta\ll Q_{\theta'}$).
Then the KLD between $P_\theta$ and $Q_{\theta'}$ was first considered in~\cite{JS-2021}:
\begin{eqnarray*}
D_\KL[P_\theta:Q_{\theta'}]&=&E_P\left[\log \left(\frac{\dP_\theta}{\dQ_{\theta'}}\right)\right],\\
&=& E_{P_\theta}\left[\left(\theta^\top t_\calP(x)- {\theta'}^\top t_\calQ(x)-F_\calP(\theta)+F_\calQ(\theta')+k_\calP(x)-k_\calQ(x)\right)
\underbrace{\frac{\dmu_\calP}{\dmu_\calQ}}_{=1}
\right],\\
&=& F_\calQ(\theta')-F_\calP(\theta)+\theta^\top E_{P_\theta}[t_\calP(x)]-{\theta'}^\top E_{P_\theta}\left[t_\calQ(x)\right]
+E_{P_\theta}\left[k_\calP(x)-k_\calQ(x)\right].
\end{eqnarray*}

Recall that the dual parameterization of an exponential family density $P_\theta$ is $P^\eta$ with $\eta=E_{P_\theta}[t_\calP(x)]=\nabla F_\calP(\theta)$~\cite{EF-2019}, and that the Fenchel-Young equality is $F(\theta)+F^*(\eta)=\theta^\top\eta$ for $\eta=\nabla F(\theta)$.
Thus the KLD between $P_\theta$ and $Q_{\theta'}$ can be rewritten as
\begin{equation}\label{eq:KLDEFs}
\boxed{D_\KL[P_\theta:Q_{\theta'}] = F_\calQ(\theta')+F_\calP^*(\eta)-{\theta'}^\top E_{P_\theta}\left[t_\calQ(x)\right]
+E_{P_\theta}\left[k_\calP(x)-k_\calQ(x)\right].}
\end{equation}

This formula was reported in~\cite{JS-2021} and generalizes the Fenchel-Young divergence~\cite{acharyya2013learning} obtained 
when $\calP=\calQ$ (with $t_\calP(x)=t_\calQ(x)$, $k_\calP(x)=k_\calQ(x)$, and $F(\theta)=F_\calP(\theta)=F_\calQ(\theta)$ and $F^*(\eta)=F_\calP^*(\eta)=F_\calQ^*(\eta)$).

The formula of Eq.~\ref{eq:KLDEFs} was illustrated in~\cite{JS-2021} with two examples:
The KLD between Laplacian distributions and zero-centered Gaussian distributions, and the KLD between two Weibull distributions.
Both these examples use the Lebesgue base measure for $\mu_\calP$ and $\mu_\calQ$.

Let us report another example which uses the counting measure as the base measure for $\mu_\calP$ and $\mu_\calQ$.

\begin{Example}
Consider the KLD between a Poisson probability mass function (pmf) and a geometric pmf.
The canonical decomposition of the Poisson and geometric pmfs are summarized in Table~\ref{tab:comparison}.
The KLD between a Poisson pmf $p_\lambda$ and a geometric pmf $q_p$ is equal to
\begin{eqnarray}
D_\KL[P_\lambda: Q_p] &=&F_\calQ(\theta')+F_\calP^*(\eta)-E_{P_{\theta}}[t_\calQ(x)]\cdot\theta' +E_{P_{\theta}}[k_\calP(x)-k_\calQ(x)],\\
&=&-\log p+\lambda\log\lambda-\lambda-\lambda\log(1-p)-E_{P_\lambda}[\log x!].
\end{eqnarray}

Since $E_{p_\lambda}[-\log x!]=-\sum_{k=0}^\infty e^{-\lambda} \frac{\lambda^k\log(k!)}{k!}$, we have
$$
D_\KL[P_\lambda: Q_p]=-\log p+\lambda\log \frac{\lambda}{1-p}-\lambda-\sum_{k=0}^\infty e^{-\lambda} \frac{\lambda^k\log(k!)}{k!}
$$

Notice that we can calculate the KLD between two geometric distributions $Q_{p_1}$ and $Q_{p_2}$ as
\begin{eqnarray*}
D_\KL[Q_{p_1}:Q_{p_2}]&=&B_{F_\calQ}(\theta(p_2):\theta(p_1)),\\
&=& F_\calQ(\theta(p_2))-F_\calQ(\theta(p_1))-(\theta(p_2)-\theta(p_1))\eta(p_1),
\end{eqnarray*}
We get:
$$
D_\KL[Q_{p_1}:Q_{p_2}]=\log\left(\frac{p_1}{p_2}\right)-\left(1-\frac{1}{p_1}\right)\log \frac{1-p_1}{1-p_2}.
$$
\end{Example}

\begin{table}
\centering
\begin{tabular}{lll}
 & Poisson family $\calP$ & Geometric family $\calQ$ \\ \hline
support & $\bbN \cup\{0\}$&   $\bbN \cup\{0\}$\\
base measure & counting measure  & counting measure\\
ordinary parameter & rate $\lambda>0$ & success probability $p\in (0,1)$ \\
pmf & $ \frac{\lambda^x}{x!} \exp(-\lambda)$  & $(1-p)^x p$\\
sufficient statistic & $t_\calP(x)=x$ & $t_\calQ(x)=x$\\
natural parameter & $\theta(\lambda)=\log\lambda$ & $\theta(p)=\log(1-p)$\\
cumulant function & $F_\calP(\theta)=\exp(\theta)$ & $F_\calQ(\theta)=-\log(1-\exp(\theta))$\\
                  & $F_\calP(\lambda)=\lambda$ & $F_\calQ(p)=-\log(p)$\\
auxiliary measure term & $k_\calP(x)=-\log x!$ & $k_\calQ(x)=0$\\
moment parameter $\eta=E[t(x)]$ & $\eta=\lambda$ & $\eta=\frac{e^\theta}{1-e^\theta}=\frac{1}{p}-1$\\
negentropy (convex conjugate) & $F^*_\calP(\eta(\lambda))=\lambda\log\lambda-\lambda$ & $F^*_\calQ(\eta(p))=\left(1-\frac{1}{p}\right)\log(1-p)+\log p$\\
($F^*(\eta)=\theta\cdot\eta-F(\theta)$)
\end{tabular}
\caption{Canonical decomposition of the Poisson and the geometric discrete exponential families.}\label{tab:comparison}
\end{table}

%%%%
\section{The duo Fenchel-Young divergence and its corresponding duo Bregman divergence}\label{sec:duoyf}
%%%%
 
\begin{figure}
\centering
\begin{tabular}{cc}
\includegraphics[width=0.49\textwidth]{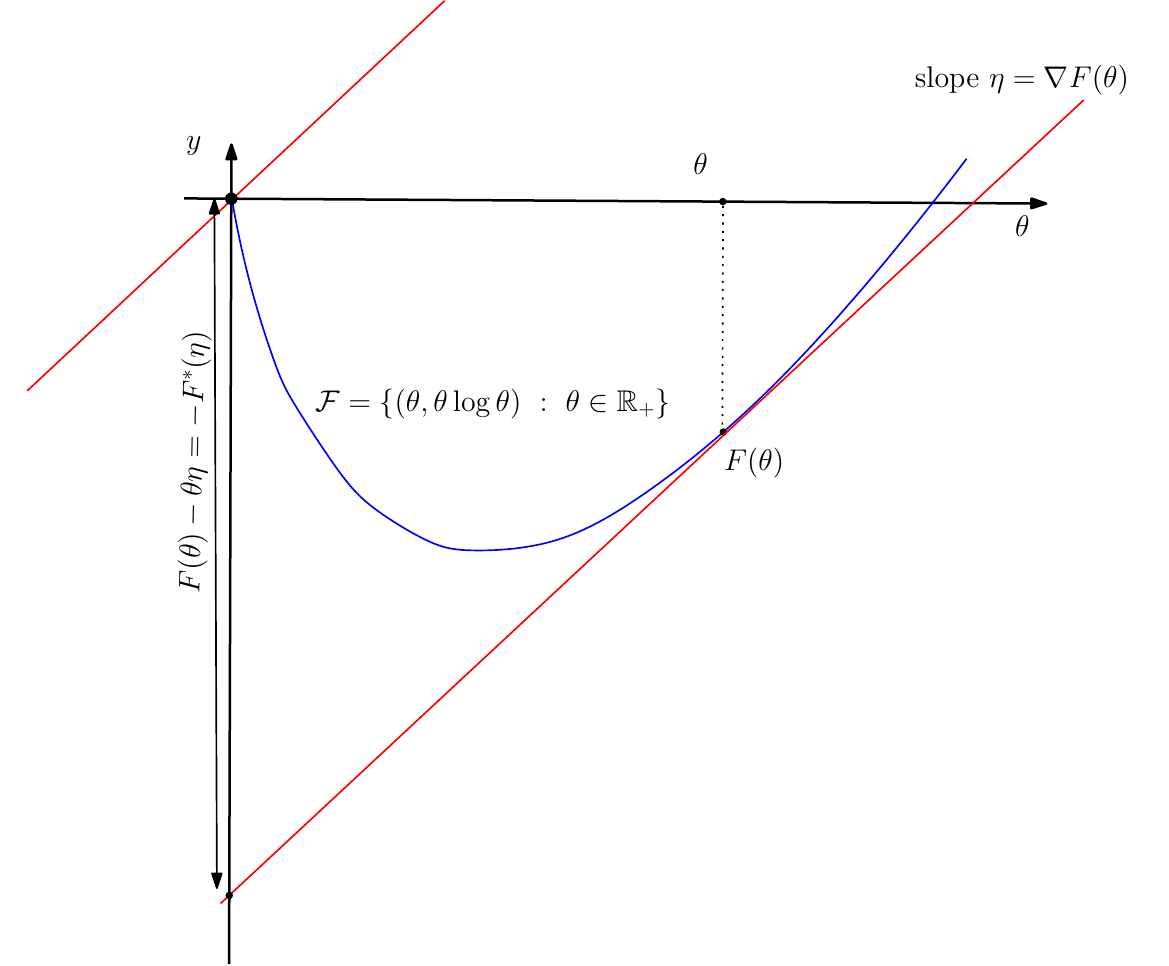}
&
\includegraphics[width=0.49\textwidth]{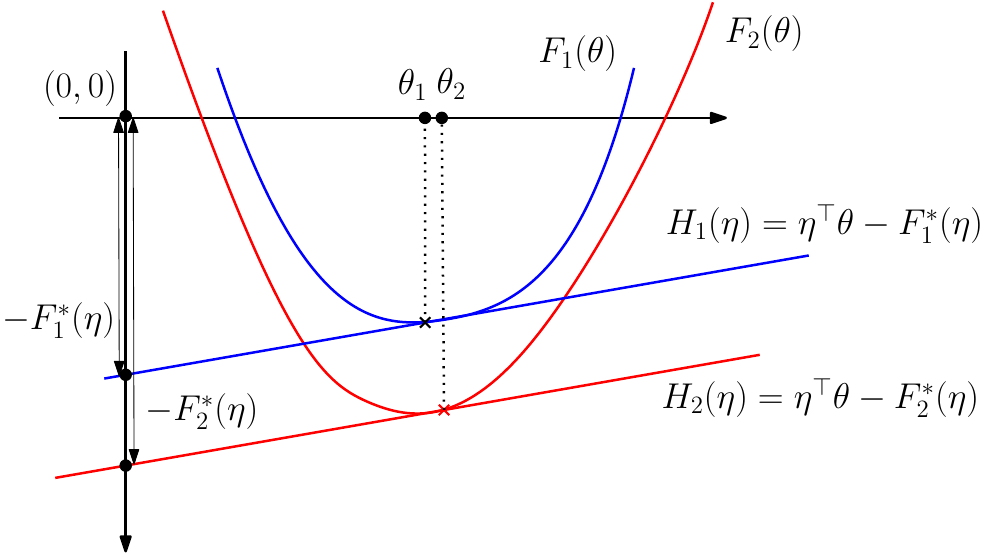}\\
(a) & (b)
\end{tabular}

\caption{(a) Visual illustration of the Legendre-Fenchel transformation: 
$F^*(\eta)$ is measured as the vertical gap between the origin and the hyperplane of ``slope'' $\eta$ tangent at $F(\theta)$ 
evaluated at $\theta=0$.
(b)
The Legendre transforms $F_1^*(\eta)$ and $F_1^*(\eta)$ of two functions $F_1(\theta)$ and $F_2(\theta)$ such that $F_1(\theta)\geq F_2(\theta)$ reverses the dominance order: $F_2^*(\eta)\geq F_1^*(\eta)$. }
\label{fig:LegendreDominance}
\end{figure}

Inspired by formula of Eq.~\ref{eq:KLDEFs}, we shall define the {\em duo Fenchel-Young divergence} using a {\em dominance condition} on a pair $(F_1(\theta),F_2(\theta))$ of strictly convex generators:

\begin{Definition}[duo Fenchel-Young divergence]\label{def:genyf}
Let  $F_1(\theta)$ and $F_2(\theta)$ be two strictly convex functions such that $F_1(\theta)\geq F_2(\theta)$ for any $\theta\in \Theta_{12}=\mathrm{dom}(F_1)\cap\mathrm{dom}(F_2)$.
Then the  duo Fenchel-Young divergence $Y_{F_1,F_2^*}(\theta,\eta')$ is defined by
\begin{equation}
\boxed{Y_{F_1,F_2^*}(\theta,\eta'):= F_1(\theta)+F_2^*(\eta')-\theta^\top\eta'.}
\end{equation}
\end{Definition}

When $F_1(\theta)=F_2(\theta)=:F(\theta)$, we have $F_1^*(\eta)=F_2^*(\eta)=:F^*(\eta)$, and we retrieve the ordinary Fenchel-Young divergence~\cite{acharyya2013learning}:
$$
Y_{F,F^*}(\theta,\eta'):=F(\theta)+F^*(\eta')-\theta^\top\eta'\geq 0.
$$

\begin{Property}[Non-negative duo Fenchel-Young divergence]
The duo Fenchel-Young divergence is always non-negative.
\end{Property}

\begin{proof}
The proof relies on the reverse dominance property of strictly convex and differentiable conjugate functions: 
Namely, if $F_1(\theta)\geq F_2(\theta)$ then we have $F_2^*(\eta)\geq F_1^*(\eta)$.
This property is graphically illustrated in Figure~\ref{fig:LegendreDominance}.
The reverse dominance property of the Legendre-Fenchel transformation can be checked algebraically as follows:
\begin{eqnarray*}
F_1^*(\eta)&=&\sup_{\theta\in\Theta} \{\eta^\top\theta-F_1(\theta)\},\\
&=& \eta^\top\theta_1-F_1(\theta_1)\quad\quad(\mbox{with $\eta=\nabla F_1(\theta_1)$}),\\
&\leq& \eta^\top\theta_1-F_2(\theta_1),\\
&\leq & \sup_{\theta\in\Theta} \{\eta^\top\theta-F_2(\theta)\}=F_2^*(\eta).
\end{eqnarray*}
Thus we have $F_1^*(\eta)\leq F_2^*(\eta)$ when $F_1(\theta)\geq F_2(\theta)$.
Therefore it follows that $Y_{F_1,F_2^*}(\theta,\eta')\geq 0$ since we have
\begin{eqnarray*}
Y_{F_1,F_2^*}(\theta,\eta') &:=& F_1(\theta)+F_2^*(\eta')-\theta^\top\eta',\\
&\geq& F_1(\theta)+F_1^*(\eta')-\theta^\top\eta' = Y_{F_1,F_1^*}(\theta,\eta')\geq 0,
\end{eqnarray*}
where $Y_{F_1,F_1^*}$ is the ordinary Fenchel-Young divergence which is guaranteed to be non-negative from the Fenchel-Young's inequality.
\end{proof}

\begin{figure}
\centering
\includegraphics[width=0.4\textwidth]{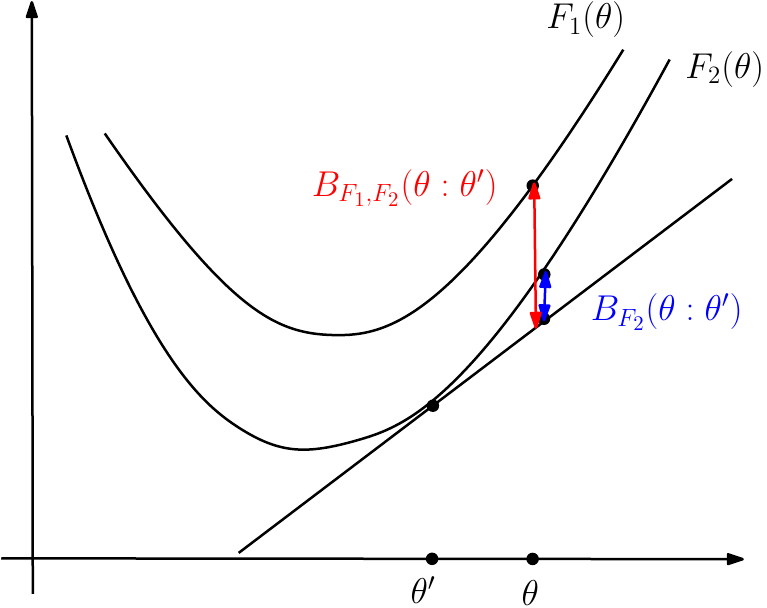}
\caption{The duo Bregman divergence induced by two strictly convex and differentiable functions $F_1$ and $F_2$ such that $F_1(\theta)\geq F_2(\theta)$: We check graphically that $B_{F_1,F_2}(\theta:\theta')\geq B_{F_2}(\theta:\theta')$ (vertical gaps).\label{fig:duoBregman}}
\end{figure}

We can express the duo Fenchel-Young divergence using the primal coordinate systems as a generalization of the Bregman divergence to two generators that we term the duo Bregman divergence (see Figure~\ref{fig:duoBregman}) :
\begin{equation}\label{eq:duobdyf}
B_{F_1,F_2}(\theta:\theta') := Y_{F_1,F_2^*}(\theta,\eta') = F_1(\theta)-F_2(\theta')-(\theta-\theta')^\top \nabla F_2(\theta'),
\end{equation}
with $\eta'=\nabla F_2(\theta')$.

This generalized Bregman divergence is non-negative when $F_1(\theta)\geq F_2(\theta)$.
Indeed, we check that
\begin{eqnarray*}
B_{F_1,F_2}(\theta:\theta')  &=&  F_1(\theta)-F_2(\theta')-(\theta-\theta')^\top \nabla F_2(\theta'),\\
&\geq& F_2(\theta)-F_2(\theta')-(\theta-\theta')^\top \nabla F_2(\theta')=B_{F_2}(\theta:\theta')\geq 0.
\end{eqnarray*}

\begin{Definition}[duo Bregman divergence]\label{def:genbd}
Let  $F_1(\theta)$ and $F_2(\theta)$ be two strictly convex functions such that $F_1(\theta)\geq F_2(\theta)$ for any $\theta\in \Theta_{12}=\mathrm{dom}(F_1)\cap\mathrm{dom}(F_2)$.
Then the  generalized Bregman divergence  is defined by
\begin{equation}
\boxed{B_{F_1,F_2}(\theta:\theta') =  F_1(\theta)-F_2(\theta')-(\theta-\theta')^\top \nabla F_2(\theta')\geq 0.}
\end{equation}
\end{Definition}

\begin{figure}
\centering
\begin{tabular}{ccc}
\includegraphics[width=0.3\textwidth]{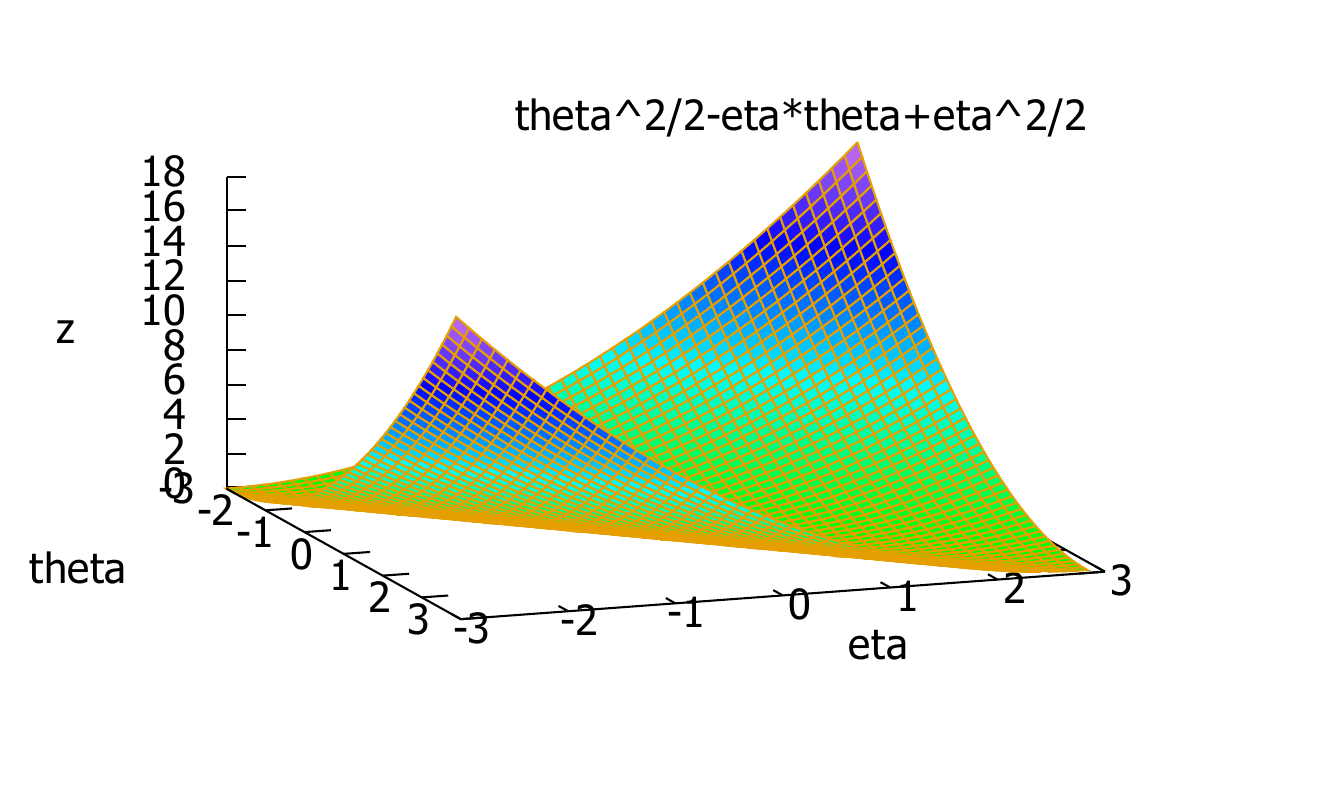}
&
\includegraphics[width=0.3\textwidth]{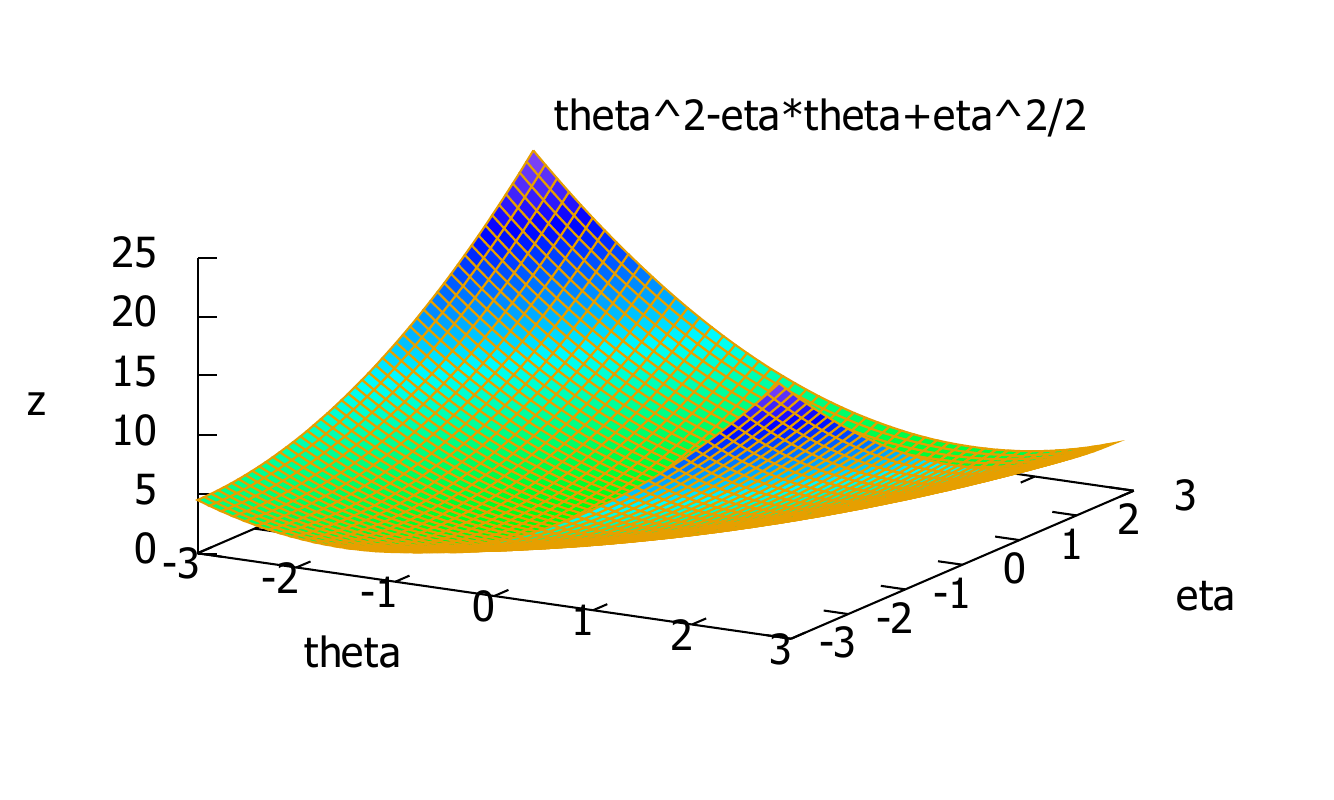}
&
\includegraphics[width=0.3\textwidth]{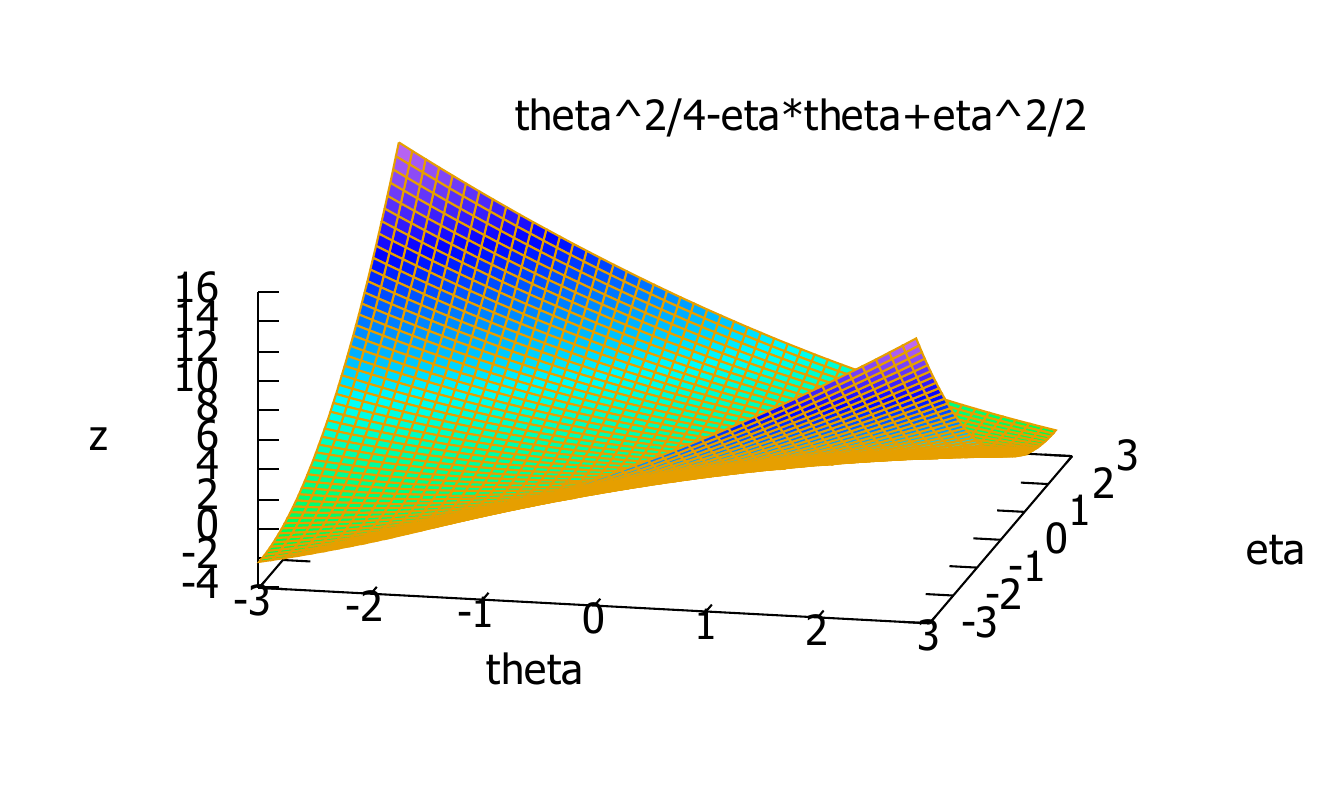}
\\
(a) & (b) & (c)
\end{tabular}

\caption{The duo half squared Euclidean distance $D_a^2(\theta:\theta'):=\frac{a}{2}\theta^2 + \frac{1}{2} {\theta'}^2 - \theta\theta'$ is non-negative when $a\geq 1$: (a) half squared Euclidean distance ($a=1$), (b) $a=2$, (c) $a=\frac{1}{2}$ which shows that the divergence can be negative then since $a<1$.}
\label{fig:exDuoSquaredEucl}
\end{figure}

\begin{Example}
Consider $F_1(\theta)=\frac{a}{2}\theta^2$ for $a>0$. We have $\eta=a\theta$, $\theta=\frac{\eta}{a}$, and
$$
F_1^*(\eta)=\frac{\eta^2}{a}-\frac{a}{2}\frac{\eta^2}{a^2}=\frac{\eta^2}{2a}.
$$
Let $F_2(\theta)=\frac{1}{2}\theta^2$ so that $F_1(\theta)\geq F_2(\theta)$ for $a\geq 1$.
We check that $F_1^*(\eta)=\frac{\eta^2}{2a}\leq F_2^*(\eta)$ when $a\geq 1$.
The duo Fenchel-Young divergence is
$$
Y_{F_1,F_2^*}(\theta,\eta') = \frac{a}{2}\theta^2 + \frac{1}{2} {\eta'}^2 - \theta\eta'\geq 0,
$$
when $a\geq 1$.
We can express the duo Fenchel-Young divergence in the primal coordinate systems as
$$
B_{F_1,F_2}(\theta,\theta)=\frac{a}{2}\theta^2 + \frac{1}{2} {\theta'}^2 - \theta\theta'.
$$
When $a=1$, $F_1(\theta)=F_2(\theta)=\frac{1}{2}\theta^2:=F(\theta)$, and 
we get $B_{F}(\theta,\theta')=\frac{1}{2}\|\theta-\theta'\|_2^2$,
half the squared Euclidean distance as expected.
Figure~\ref{fig:exDuoSquaredEucl} displays the graph plot of the duo Bregman divergence for several values $a$.
\end{Example}

\begin{figure}
\centering
\begin{tabular}{cc}
\includegraphics[width=0.49\textwidth]{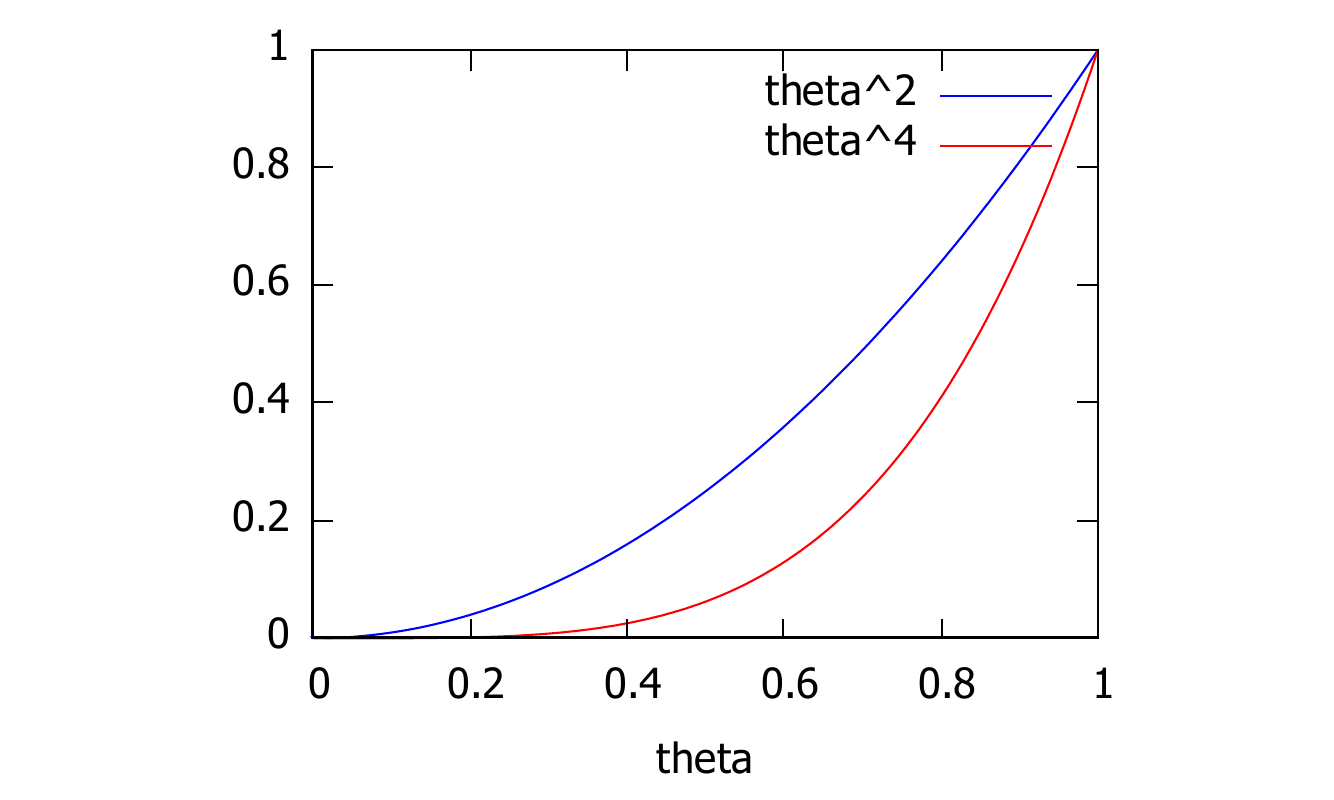}
&
\includegraphics[width=0.49\textwidth]{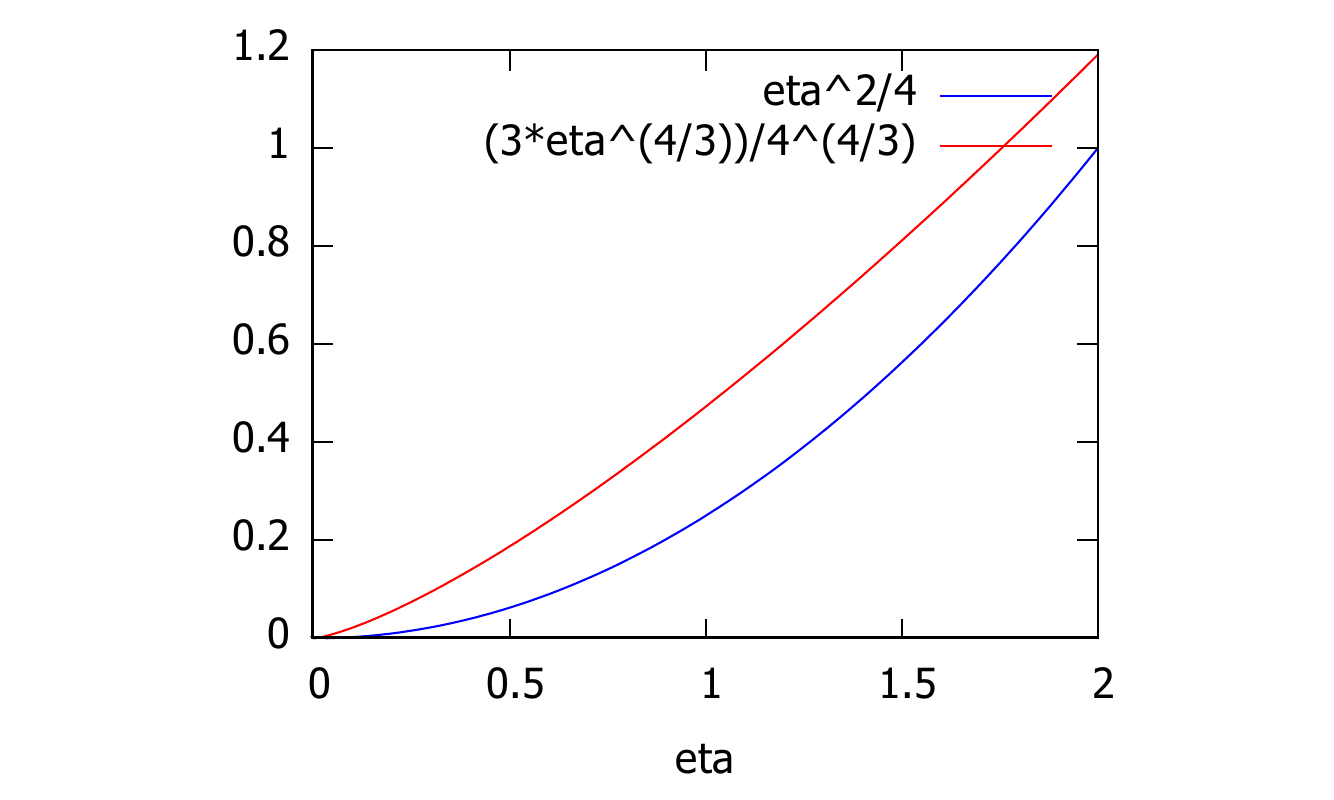}\\
Convex functions  & Conjugate functions
\end{tabular}

\caption{\label{fig:plot}The Legendre transform reverses the dominance ordering: 
$F_1(\theta)=\theta^2\geq F_2(\theta)=\theta^4 \Leftrightarrow  F_1^*(\eta)\geq F_2^*(\eta)$ for $\theta\in [0,1]$. }
\end{figure}

\begin{Example}
Consider $F_1(\theta)=\theta^2$ and $F_2(\theta)=\theta^4$ on the domain $\Theta=[0,1]$.
We have $F_1(\theta)\geq F_2(\theta)$ for $\theta\in\Theta$. 
The convex conjugate of $F_1(\eta)$ is $F_1^*(\eta)=\frac{1}{4}\eta^2$.
We have 
$$
F_2^*(\eta)={\eta}^{\frac{4}{3}} \left(\left(\frac{1}{4}\right)^{\frac{1}{3}}-\left(\frac{1}{4}\right)^{\frac{4}{3}}\right)
=\frac{3}{4^{\frac{4}{3}}}{\eta}^{\frac{4}{3}}
$$
 with $\eta_2(\theta)=4{\theta}^3$.
Figure~\ref{fig:plot} plots the convex functions $F_1(\theta)$ and $F_2(\theta)$, and their convex conjugates $F_1^*(\eta)$ and $F_2^*(\eta)$.
We observe that $F_1(\theta)\geq F_2(\theta)$ on $\theta\in [0,1]$ and that $F_1^*(\eta)\leq F_2^*(\eta)$ on $H=[0,2]$.
\end{Example}

We now state a property between dual duo Bregman divergences:
\begin{Property}[Dual duo Fenchel-Young/Bregman divergences]
We have
$$
Y_{F_1,F_2^*}(\theta:\eta')=B_{F_1,F_2}(\theta:\theta')=B_{F_2^*,F_1^*}(\eta':\eta)=Y_{F_2^*,F_1}(\eta':\theta)
$$
\end{Property}

\begin{proof}
From the Fenchel-Young equalities of the inequalities, 
we have $F_1(\theta)=\theta^\top\eta-F_1^*(\eta)$ for $\eta=\nabla F_1(\theta)$ and
$F_2(\theta')={\theta'}^\top \eta'-F_2^*(\eta')$ with $\eta'=\nabla F_2(\theta')$.
Thus we have
\begin{eqnarray*}
B_{F_1,F_2}(\theta:\theta') &=&  F_1(\theta)-F_2(\theta')-(\theta-\theta')^\top \nabla F_2(\theta'),\\
&=& \theta^\top\eta-F_1^*(\eta) -{\theta'}^\top \eta'+F_2^*(\eta')-(\theta-\theta')^\top \eta',\\
&=& F_2^*(\eta')-F_1^*(\eta)-(\eta'-\eta)^\top\theta,\\
&=& B_{F_2^*,F_1^*}(\eta':\eta).
\end{eqnarray*}
Recall that $F_1(\theta)\geq F_2(\theta)$ implies that $F_1^*(\eta)\leq F_2^*(\eta)$, and therefore the dual duo Bregman divergence is non-negative: $ B_{F_2^*,F_1^*}(\eta':\eta)\geq 0$.
\end{proof}

%%%%%%%%%
\section{Signed duo Bregman pseudo-divergences and DC programming}
%%%%%%%%%

Consider the following optimization problem:
$$
\min_{\theta\in\Theta} F(\theta),\quad F(\theta)=F_1(\theta)-F_2(\theta),
$$
where both $F_1$ and $F_2$ are strictly convex and differentiable convex functions of Legendre type~\cite{LegendreType-1967}.
Since $F$ is the difference of convex (DC) functions, this problem has been called DC programming~\cite{horst1999dc,an2005dc} in the literature.

In the above case  both strictly convex and smooth $F_1$ and $F_2$ functions, the DC Algorithm (DCA)~\cite{an2005dc} is the following simple method to solve iteratively this minimization problem:
Start with $\theta_0\in\Theta$, and then iteratively update as follows:
$$
\theta_{t+1}=(\nabla F_1)^{-1}(\nabla F(\theta_t)).
$$
The DCA decreases monotonically $F$ and converges to a minimum or a saddle point.
Convergence analysis of the DCA has been studied for the general case of  extended convex functions $F_1$ and $F_2$ in~\cite{ConvergenceDCA-2024}.
The DCA has also been called Convex-ConCave Procedure~\cite{yuille2001concave,CCCP-FrankWolfe-2022}.

The DCA can be derived by considering a first-order Taylor expansion of $F_2$ at $\theta_t$ as follows:
\begin{eqnarray*}
\theta_{t+1} &=& \arg\min_\theta F_1(\theta)-F_2(\theta_t)-(\theta-\theta_t)^\top \nabla F_2(\theta_t),\\
&=&  \arg\min_\theta B_{F_1,F_2}(\theta:\theta_t),
\end{eqnarray*}
where $B_{F_1,F_2}$ is a signed duo Bregman pseudo-divergence:
That is, we relax  $F_1$ and $F_2$ to arbitrary strictly convex and differentiable generators and do not enforce $F_1\geq F_2$ so that the duo Bregman pseudo-divergence may be negative.

%%%%
\section{Kullback-Leibler divergence between a truncated density and a density of an exponential family}\label{sec:kldtrunc}
%%%
Let $\calE_1=\{P_\theta\st \theta\in\Theta_1\}$ be an exponential family of distributions all dominated by $\mu$ with Radon-Nikodym density
$p_\theta(x)=\exp(\theta^\top t(x)-F_1(\theta)+k(x))\,\dmu(x)$ defined on the support $\calX_1$.
Let $\calE_2=\{Q_\theta\st \theta\in\Theta_2\}$ be another exponential family of distributions all dominated by $\mu$ with Radon-Nikodym density
$q_\theta(x)=\exp(\theta^\top t(x)-F_2(\theta)+k(x))\,\dmu(x)$ defined on the support $\calX_2$ such that $\calX_1\subseteq \calX_2$.
Let $\tilde{p}_\theta(x)=\exp(\theta^\top t(x)+k(x))\,\dmu(x)$ be the common unnormalized density so that
$$
p_\theta(x)=\frac{\tilde{p}_\theta(x)}{Z_1(\theta)}
$$
and
$$
q_\theta(x)=\frac{\tilde{p}_\theta(x)}{Z_2(\theta)}=\frac{Z_1(\theta)}{Z_2(\theta)}\, p_\theta(x),
$$
with $Z_1(\theta)=\exp(F_1(\theta))$ and $Z_2(\theta)=\exp(F_2(\theta))$ are the log-normalizer functions of $\calE_1$ and $\calE_2$, respectively.

We have
\begin{eqnarray*}
D_\KL[p_{\theta_1}:q_{\theta_2}] &=&\int_{\calX_1} p_{\theta_1}(x)\log \frac{p_{\theta_1}(x)}{q_{\theta_2}(x)}\dmu(x),\\
&=& \int_{\calX_1} p_{\theta_1}(x)\log \frac{p_{\theta_1}(x)}{p_{\theta_2}(x)} \dmu(x)+
 \int_{\calX_1} p_{\theta_1}(x)\log \left( \frac{Z_2(\theta_2)}{Z_1(\theta_2)}\right),\\
&=& D_\KL[p_{\theta_1}:p_{\theta_2}]+\log  Z_2(\theta_2)-\log  Z_1(\theta_2).
\end{eqnarray*}

Since $D_\KL[p_{\theta_1}:p_{\theta_2}]=B_{F_1}(\theta_2:\theta_1)$ and $\log Z_i(\theta)=F_i(\theta)$, we get
\begin{eqnarray*}
D_\KL[p_{\theta_1}:q_{\theta_2}] &=& B_{F_1}(\theta_2:\theta_1) + F_2(\theta_2)-F_1(\theta_2),\\
&=& F_1(\theta_2)-F_1(\theta_1)-(\theta_2-\theta_1)^\top \nabla F_1(\theta_1) + F_2(\theta_2)-F_1(\theta_2),\\
&=& F_2(\theta_2)-F_1(\theta_1)-(\theta_2-\theta_1)^\top \nabla F_1(\theta_1)  =: B_{F_2,F_1}(\theta_2:\theta_1).
\end{eqnarray*}

Observe that we have:
$$
F_2(\theta)=
\log\int_{\calX_2} \tilde{p}_\theta(x)\, \dmu(x)\geq 
\log\int_{\calX_1} \tilde{p}_\theta(x)))\,\dmu(x):=F_1(\theta),
$$
since $\calX_1\subseteq\calX2$. Therefore $\Theta_2\subseteq\Theta_1$, and the common natural parameter space is 
$\Theta_{12}=\Theta_1\cap \Theta_2=\Theta_2$.

Notice that the reverse Kullback-Leibler divergence $D_\KL^*[p_{\theta_1}:q_{\theta_2}]=D_\KL[q_{\theta_2}:p_{\theta_1}]=+\infty$ since $Q_{\theta_2}\not\ll P_{\theta_1}$.

\begin{Theorem}[Kullback-Leibler divergence between nested exponential family densities]\label{thm:KLDnestedEF}
Let $\calE_2=\{q_{\theta_2}\}$ be an exponential family with support $\calX_2$, and $\calE_1=\{p_{\theta_1}\}$ a truncated exponential family of $\calE_2$ with support $\calX_1\subset\calX_2$. Let $F_1$ and $F_2$ denote the log-normalizers of $\calE_1$ and $\calE_2$ and $\eta_1$ and $\eta_2$ the moment parameters corresponding to the natural parameters $\theta_1$ and $\theta_2$.
Then the Kullback-Leibler divergence between a truncated density of $\calE_1$ and a
density of $\calE_2$ is
$$
D_\KL[p_{\theta_1}:q_{\theta_2}]=Y_{F_2,F_1^*}(\theta_2:\eta_1)=B_{F_2,F_1}(\theta_2:\theta_1)=B_{F_1^*,F_2^*}(\eta_1:\eta_2)=Y_{F_1^*,F_2}(\eta_1:\theta_2).
$$
\end{Theorem}

For example, consider the calculation of the KLD between an exponential distribution (view as half a Laplacian distribution) and a Laplacian distribution.

\begin{Example}
Let $\calE_1=\{p_\lambda(x)=\lambda \exp(-\lambda x), \lambda\in\bbR_{++}, x>0 \}$ and 
$\calE_2=\{q_\lambda(x)=\lambda \exp(-\lambda |x|), \lambda\in\bbR_{++}, x\in\bbR \}$ denote the exponential families of exponential distributions and Laplacian distributions, respectively.
We have the sufficient statistic $t(x)=-|x|$ and natural parameter $\theta=\lambda$ so that
$\tilde{p}_\theta(x)=\exp(-|x|\theta)$.
The log-normalizers are $F_1(\theta)=-\log\theta$ and $F_2(\theta)=-\log\theta+\log 2$ (hence $F_2(\theta)\geq F_1(\theta)$).
The moment parameter $\eta=\nabla F_1(\theta)=\nabla F_2(\theta)=-\frac{1}{\theta}=-\frac{1}{\lambda}$.
Thus using the duo Bregman divergence, we have:
\begin{eqnarray*}
D_\KL[p_{\theta_1}:q_{\theta_2}] &=& B_{F_2,F_1}(\theta_2:\theta_1),\\
&=& F_2(\theta_2)-F_1(\theta_1)-(\theta_2-\theta_1)^\top \nabla F_1(\theta_1),\\
&=& \log 2+\log \frac{\lambda_1}{\lambda_2}+\frac{\lambda_2}{\lambda_1}-1.
\end{eqnarray*}
Moreover, we can interpret that divergence using the Itakura-Saito divergence~\cite{itakura1968analysis}:
$$
D_\IS[\lambda_1:\lambda_2]:=\frac{\lambda_1}{\lambda_2}-\log\frac{\lambda_1}{\lambda_2}-1\geq 0.
$$
we have
$$
D_\KL[p_{\theta_1}:q_{\theta_2}]=D_\IS[\lambda_2:\lambda_1]  +\log 2\geq 0.
$$

We check the result using the duo Fenchel-Young divergence:
$$
D_\KL[p_{\theta_1}:q_{\theta_2}] =Y_{F_2,F_1^*}(\theta_2:\eta_1),
$$
with $F_1^*(\eta)=-1+\log \left(-\frac{1}{\eta}\right)$:
\begin{eqnarray*}
D_\KL[p_{\theta_1}:q_{\theta_2}] &=& Y_{F_2,F_1^*}(\theta_2:\eta_1),\\
&=& -\log\lambda_2+\log 2 -1+\log \lambda_1 +\frac{\lambda_2}{\lambda_1},\\
&=& \log \frac{\lambda_1}{\lambda_2}+\frac{\lambda_2}{\lambda_1}+\log_2 -1.
\end{eqnarray*}
\end{Example}

Next, consider the calculation of the KLD between a half-normal distribution and a (full) normal distribution:

\begin{Example}
Consider $\calE_1$ and $\calE_2$ be the scale family of the half standard normal distributions and the scale family of the standard normal distribution, respectively.
We have $\tilde{p}_\theta(x)=\exp\left(-\frac{x^2}{2\sigma^2}\right)$ with 
$Z_1(\theta)=\sigma\sqrt{\frac{\pi}{2}}$ and $Z_2(\theta)=\sigma\sqrt{2\pi}$.
Let the sufficient statistic be $t(x)=-\frac{x^2}{2}$ so that the natural parameter is $\theta=\frac{1}{\sigma^2}\in \bbR_{++}$.
Here, we have both $\Theta_1=\Theta_2= \bbR_{++}$.
For this example, we check that $Z_1(\theta)=\frac{1}{2}\, Z_2(\theta)$.
We have $F_1(\theta)=-\frac{1}{2}\log\theta+\frac{1}{2}\log\frac{\pi}{2}$
and $F_2(\theta)=-\frac{1}{2}\log\theta+\frac{1}{2}\log(2\pi)$ (with $F_2(\theta)\geq F_1(\theta)$).
We have $\eta=-\frac{1}{2\theta}=-\frac{1}{2}\sigma^2$.
The KLD between two half scale normal distributions is
\begin{eqnarray*}
D_\KL[p_{\theta_1}:p_{\theta_2}] &=& B_{F_1}(\theta_2:\theta_1),\\
&=& \frac{1}{2}\left(\log\frac{\sigma_2^2}{\sigma_1^2}+\frac{\sigma_1^2}{\sigma_2^2}-1\right).
\end{eqnarray*}
Since $F_1(\theta)$ and $F_2(\theta)$ differ only by a constant and that the Bregman divergence is invariant under an affine term of its generator, we have
\begin{eqnarray*}
D_\KL[q_{\theta_1}:q_{\theta_2}] &=& B_{F_2}(\theta_2:\theta_1),\\
&=& B_{F_1}(\theta_2:\theta_1)= \frac{1}{2}\left(\log\frac{\sigma_2^2}{\sigma_1^2}+\frac{\sigma_1^2}{\sigma_2^2}-1\right).
\end{eqnarray*}
Moreover, we can interpret those Bregman divergences as half of the Itakura-Saito divergence:
$$
D_\KL[p_{\theta_1}:p_{\theta_2}]=D_\KL[q_{\theta_1}:q_{\theta_2}]=B_{F_2}(\theta_2:\theta_1)=\frac{1}{2}D_\IS[\sigma_1^2:\sigma_2^2],
$$
where
$$
D_\IS[\lambda_1:\lambda_2]:=\frac{\lambda_1}{\lambda_2}-\log\frac{\lambda_1}{\lambda_2}-1.
$$

It follows that
\begin{eqnarray*}
D_\KL[p_{\theta_1}:q_{\theta_2}] &=& B_{F_2,F_1}(\theta_2:\theta_1) = F_2(\theta_2)-F_1(\theta_1)-(\theta_2-\theta_1)^\top \nabla F_1(\theta_1),\\
&=& \frac{1}{2} \left( \log\frac{\sigma_2^2}{\sigma_1^2}+{\sigma_1^2}{\sigma_2^2}+\log 4-1 \right).
\end{eqnarray*}
Notice that $\frac{1}{2}\log 4=\log 2>0$ so that $D_\KL[p_{\theta_1}:q_{\theta_2}]\geq D_\KL[q_{\theta_1}:q_{\theta_2}]$.
\end{Example}

Thus the Kullback-Leibler divergence between a truncated density and another density of the same exponential family amounts to calculate a duo Bregman divergence on the reverse parameter order: $D_\KL[p_{\theta_1}:q_{\theta_2}]=B_{F_2,F_1}(\theta_2:\theta_1)$.
Let $D_\KL^*[p:q]:=D_\KL[q:p]$ be the reverse Kullback-Leibler divergence. 
Then $D_\KL^*[q_{\theta_2}:p_{\theta_1}]=B_{F_2,F_1}(\theta_2:\theta_1)$.

Notice that truncated exponential families are also exponential families but they may be non-steep~\cite{del1994singly}.

The next example shows how to compute the Kullback-Leibler divergence between two truncated normal distributions:

% https://en.wikipedia.org/wiki/Truncated_normal_distribution
\begin{Example}\label{ex:truncnormal}
Let $N_{a,b}(m,s)$ denote a truncated normal distribution with support the open interval $(a,b)$ ($a<b$) and probability density function (pdf) defined by:
$$
p_{m,s}^{a,b}(x)=\frac{1}{Z_{a,b}(m,s)}\, \exp\left(-\frac{(x-m)^2}{2s^2}\right),
$$
where $Z_{a,b}(m,s)$ is related to the partition function~\cite{truncatednormal-2014} expressed using the cumulative distribution function (CDF) $\Phi_{m,s}(x)$:
$$
Z_{a,b}(m,s)=\sqrt{2\pi}s\,\left(\Phi_{m,s}(b)-\Phi_{m,s}(a)\right),
$$
with
$$
\Phi_{m,s}(x)=\frac{1}{2} \left(1+\erf\left(\frac{x-m}{\sqrt{2}s}\right)\right),
$$
where $\erf(x)$ is the error function:
$$
\erf(x):=\frac{2}{\sqrt{\pi}}\,\int_0^x e^{-t^2} \dt.
$$
Thus we have $\erf(x)=2\,\Phi(\sqrt{2}x)-1$ where $\Phi(x)=\Phi_{0,1}(x)$.

The pdf can also be written as
$$
p_{m,s}^{a,b}(x)= \frac{1}{\sigma} \frac{\phi(\frac{x-m}{s})}{\Phi(\frac{b-m}{s})-\Phi(\frac{a-m}{s})},
$$
where $\phi(x)$ denotes the standard normal pdf ($\phi(x)=p_{0,1}^{-\infty,+\infty}(x)$):
$$
\phi(x):=\frac{1}{\sqrt{2\pi}}\, \exp\left(-\frac{x^2}{2}\right),
$$
and  $\Phi(x)=\Phi_{0,1}(x)=\int_{-\infty}^x \phi(t)\, \dt$ is the standard normal CDF.
When $a=-\infty$ and $b=+\infty$,  we have $Z_{-\infty,\infty}(m,s)=\sqrt{2\pi}\, s$ since $\Phi(-\infty)=0$ and $\Phi(+\infty)=1$.

The density $p_{m,s}^{a,b}(x)$ belongs to an exponential family $\calE_{a,b}$ with  natural parameter 
$\theta=\left(\frac{m}{s^2},-\frac{1}{2s^2}\right)$, sufficient statistics $t(x)=(x,x^2)$, and log-normalizer:
$$
F_{a,b}(\theta)=-\frac{\theta_1^2}{4\theta_2}+\log Z_{a,b}(\theta)
$$
The natural parameter space is $\Theta=\bbR\times\bbR_{--}$ where $\bbR_{--}$ denotes the set of negative real numbers.

The log-normalizer can be expressed using the source parameters $(m,s)$ (which are not the mean and standard deviation when the support is truncated, hence the notation $m$ and $s$):
\begin{eqnarray*}
F_{a,b}(m,s)&=& \frac{m^2}{2s^2}+\log Z_{a,b}(m,s),\\
&=& \frac{m^2}{2s^2}+\frac{1}{2}\log 2\pi s^2+\log \left(\Phi_{m,s}(b)-\Phi_{m,s}(a)\right).
\end{eqnarray*}

We shall use the fact that the gradient of the log-normalizer of any exponential family distribution amounts to the expectation of the sufficient statistics~\cite{EF-2019}:
$$
\nabla F_{a,b}(\theta)=E_{p_{m,s}^{a,b}}[t(x)]=\eta.
$$
Parameter $\eta$ is called the moment or expectation parameter~\cite{EF-2019}.

The mean $\mu(m,s;a,b)=E_{p_{m,s}^{a,b}}[x]=\frac{\partial}{\partial\theta_1}F_{a,b}(\theta)$ and 
 the variance $\sigma^2(m,s;a,b)=E_{p_{m,s}^{a,b}}[x^2]-\mu^2$  (with $E_{p_{m,s}^{a,b}}[x^2]=\frac{\partial}{\partial\theta_2}F_{a,b}(\theta)$)  of the truncated normal $p_{m,s}^{a,b}$ 
can be expressed using the following formula~\cite{kotz1994balakrishan,truncatednormal-2014} (page 25):
\begin{eqnarray*}
\mu(m,s;a,b) &=&m-s \, \frac{\phi(\beta)-\phi(\alpha)}{\Phi(\beta)-\Phi(\alpha)},\\
\sigma^{2}(m,s;a,b) &=&s^2 \, \left(1-\frac{\beta \phi(\beta)-\alpha \phi(\alpha)}{\Phi(\beta)-\Phi(\alpha)}-\left(\frac{\phi(\beta)-\phi(\alpha)}{\Phi(\beta)-\Phi(\alpha)}\right)^{2}\right),
\end{eqnarray*}
where $\alpha:=\frac{a-m}{s}$ and $\beta:=\frac{b-m}{s}$.
Thus we have the following moment parameter $\eta=(\eta_1,\eta_2)$ with
\begin{eqnarray}
\eta_1(m,s;a,b)&=&E_{p_{m,s}^{a,b}}[x]=\mu(m,s;a,b),\\
\eta_2(m,s;a,b)&=&E_{p_{m,s}^{a,b}}[x^2]=\sigma^2(m,s;a,b)+\mu^2(m,s;a,b).
\end{eqnarray}

Now consider two truncated normal distributions $p_{m_1,s_1}^{a_1,b_1}$ 
and $p_{m_2,s_2}^{a_2,b_2}$ with $[a_1,b_1]\subseteq [a_2,b_2]$ (otherwise, we have $D_\KL[p_{m_1,s_1}^{a_1,b_1}:p_{m_2,s_2}^{a_2,b_2}]=+\infty$).
Then the KLD between $p_{m_1,s_1}^{a_1,b_1}$ and $p_{m_2,s_2}^{a_2,b_2}$ is equivalent to a duo Bregman divergence:
\begin{eqnarray}
D_\KL[p_{m_1,s_1}^{a_1,b_1}:p_{m_2,s_2}^{a_2,b_2}] &=& F_{m_2,s_2}(\theta_2)-F_{m_1,s_1}(\theta_1)-(\theta_2-\theta_1)^\top \nabla F_{m_1,s_1}(\theta_1),\nonumber\\
&=& \frac{m_2^2}{2s_2^2}-\frac{m_1^2}{2s_1^2}+\log \frac{Z_{a_2,b_2}(m_2,s_2)}{Z_{a_1,b_1}(m_1,s_1)}
-\left(\frac{m_2}{s_2^2}- \frac{m_1}{s_1^2}\right)\eta_1(m_1,s_1;a_1,b_1)\nonumber\\
&& -\left(\frac{1}{2s_1^2}-\frac{1}{2s_2^2}\right)\eta_2(m_1,s_1;a_1,b_1).
\end{eqnarray}
Notice that $F_{m_2,s_2}(\theta) \geq F_{m_1,s_1}(\theta)$.

The code \url{https://franknielsen.github.io/IG/KLDTruncatedNormalDistributions.java} implements this  KLD between truncated univariate normal distributions.

This formula is valid for (1) the KLD between two nested truncated normal distributions, or 
for (2) the KLD between a truncated normal distribution and a (full support) normal distribution.
Notice that formula depends on the erf function used in function $\Phi$.
Furthermore, when $a_1=a_2=-\infty$ and $b_1=b_2=+\infty$, we recover (3) the KLD between two univariate normal distributions since 
$\log \frac{Z_{a_2,b_2}(m_2,s_2)}{Z_{a_1,b_1}(m_1,s_1)}=\log\frac{\sigma_2}{\sigma_1}=\frac{1}{2}\log\frac{\sigma_2^2}{\sigma_1^2}$:

%0.5*((v1/v2)+Math.log(v2/v1)+((m1-m2)*(m1-m2)/v2)-1.0);

$$
D_\KL[p_{m_1,s_1}:p_{m_2,s_2}]=\frac{1}{2}\,\left(
\log\frac{s_2^2}{s_1^2}+\frac{\sigma_1^2}{\sigma_2^2} +\frac{(m_2-m_1)^2}{s_2^2}-1.
\right).
$$

Notice that for full support normal distributions, $\mu(m,s;-\infty;+\infty)=m$ and $\sigma^2(m,s;-\infty;+\infty)=s^2$.

The entropy of a truncated normal distribution (an exponential family~\cite{nielsen2010entropies}) is 
$h[p_{m,s}^{a,b}]=-\int_a^b p_{m,s}^{a,b}(x)\log p_{m,s}^{a,b}\dx=-F^*(\eta)=F(\theta)-\theta^\top\eta$.
We find that
$$
h[p_{m,s}^{a,b}]=\log\left(\sqrt{2\pi e}s\, \left(\Phi(\beta)-\Phi(\alpha)\right)\right)+\frac{\alpha\phi(\alpha)-\beta\phi(\beta)}{2\left(\Phi(\beta)-\Phi(\alpha)\right)}.
$$
When $(a,b)=(-\infty,\infty)$, we have $\Phi(\beta)-\Phi(\alpha)=1$ and 
$\alpha\phi(\alpha)-\beta\phi(\beta)=0$ since 
$\beta=-\alpha$, $\phi(-x)=\phi(x)$ (an even function), and $\lim_{\beta\rightarrow +\infty} \beta\phi(\beta)=0$.
Thus we recover the differential entropy of a normal distribution: $h[p_{\mu,\sigma}]=\log\left(\sqrt{2\pi e}\sigma\right)$.
\end{Example}

\def\tp{{\tilde p}}
\def\tq{{\tilde q}}

Consider the Kullback-Leibler divergence extended to positive densities as follows:
$$
D_\KL[\tp:\tq]=\int\left(\tp(x)\log\frac{\tp(x)}{\tq(x)}+\tq(x)-\tp(x)\right)\,\dmu(x),
$$
where $\tp(x)$ and $\tq(x)$ are positive densities, not necessarily normalized to one.

When $\tp(x)=\exp(\inner{\theta_1}{t(x)})=\tp_{\theta_1}(x)$ and $\tq(x)=\exp(\inner{\theta_2}{t(x)})=\tp_{\theta_2}(x)$ are two unnormalized densities of a same exponential family 
$$
\left\{p_\theta(x)=\exp(\inner{t(x)}{\theta}-F(\theta))=\frac{1}{Z(\theta)}\tp_\theta(x), \quad x\in\calX,\theta\in\Theta\right\},
$$ 
with $F(\theta)=\log Z(\theta)$ and $\tp_\theta(x)$ the unnormalized densities, we have
$$
D_\KL[\tp_{\theta_1}:\tp_{\theta_2}]=B_Z(\theta_2:\theta_1),
$$
where $Z(\theta)$ is the partition function (log-convex function hence convex~\cite{nielsen2024divergences}).

Furthermore, if we truncate $\tp_{\theta_1}'$ to support $\calX_1$ and $\tp_{\theta_2}'$ to support $\calX_2$ with $\calX_1\subset\calX_2\subset\calX$, then we have
\begin{eqnarray*}
D_\KL[\tp_{\theta_1}':\tp_{\theta_2}'] &=& F_2(\theta_2)-F_1(\theta_1)-\inner{\theta_2-\theta_1}{\nabla Z_2(\theta_2)},\\
&=& B_{Z_2,Z_1}(\theta_2:\theta_1),
\end{eqnarray*}
where $Z_1(\theta)=\int_{\calX_1}\exp(\inner{t(x)}{\theta})\dmu(x)$ and $Z_2(\theta)=\int_{\calX_2}\exp(\inner{t(x)}{\theta})\dmu(x)\geq Z_1(\theta1)$.

That is, the extended KLD between two truncated unnormalized exponential family densities with nested support amounts to a duo Bregman divergence with respect to convex partition function generators $Z_1$ and $Z_2$.

\begin{Example}
Consider the unnormalized exponential distributions with $\tp_\theta(x)=\exp(-\theta x)$ with sufficient statistics $t(x)=-x$ and natural parameter $\theta>0$.
The partition function is $Z(\theta)=\frac{1}{\theta}$ which is strictly convex on $\theta>0$.
We have $Z_{[a,b]}(\theta)=\frac{1}{\theta}\frac{\exp(b\theta)-\exp(a\theta)}{\exp((b-a)\theta)}$. Let $\calX_1=[a_1,b_1]\subset\calX_2=[a_2,b_2]$ and $\tp_{\theta_1}'$ and $\tp_{\theta_2}'$ be the truncated exponential distributions on $\calX_1$ and $\calX_2$, respectively.
We have $D_\KL[\tp_{\theta_1}':\tp_{\theta_2}'] =   B_{Z_2,Z_1}(\theta_2:\theta_1)$:\\
\noindent\scalebox{0.65}{$D_\KL[\tp_{\theta_1}':\tp_{\theta_2}'] ={  {{\left(\left(\left(\left(\left(a_{1}\,\theta_{1}+1\right)\,e^{
 b_{1}\,\theta_{1}}+\left(-b_{1}\,\theta_{1}-1\right)\,e^{a_{1}
 \,\theta_{1}}\right)\,\theta_{2}^2+\left(\left(-a_{1}\,
 \theta_{1}^2-2\,\theta_{1}\right)\,e^{b_{1}\,\theta_{1}}+
 \left(b_{1}\,\theta_{1}^2+2\,\theta_{1}\right)\,e^{a_{1}\,
 \theta_{1}}\right)\,\theta_{2}\right)\,e^{a_{1}\,\theta_{2}
 }+\theta_{1}^2\,e^{b_{1}\,\theta_{1}+a_{1}\,\theta_{1}}
 \right)\,e^{b_{1}\,\theta_{2}}-\theta_{1}^2\,e^{a_{1}\,
 \theta_{2}+b_{1}\,\theta_{1}+a_{1}\,\theta_{1}}\right)\,e^{
 -b_{1}\,\theta_{2}-a_{1}\,\theta_{2}-b_{1}\,\theta_{1}-
 a_{1}\,\theta_{1}}}\over{\theta_{1}^2\,\theta_{2}}} }$
}
\end{Example}

Similarly, we have $D_\KL[\tp_{\theta_1}:\tp_{\theta_2}]=B_{F_2,F_1}(\theta_2:\theta_1)$ where 
$F_1(\theta)=\log \int_{\calX_1}\exp(\inner{t(x)}{\theta})\dmu(x)=\log Z_1(\theta)$ and $F_2(\theta)=\log Z_2(\theta)=\log\int_{\calX_2}\exp(\inner{t(x)}{\theta})\dmu(x)\geq F_1(\theta1)$.

Last, we may define a KLD-type pseudo-divergence between three densities as follows: 
$$
D_\KL[p:q|r]:=\int r(x)\log\frac{p(x)}{q(x)} \dmu(x)=E_r\left[\log\frac{p(x)}{q(x)}\right].
$$
Notice that depending on the choice of density $r$ this $3$-density KLD generalization may potentially be negative.
This generalization is sometimes met in statistics and information theory~\cite{grunwald2007minimum}.

When the densities $p=p_{\theta_1}, q=p_{\theta_2}, r=p_\theta$ all belong to the same exponential family, we have
\begin{eqnarray}
D_\KL[p_{\theta_1}:p_{\theta_2}|p_\theta] &=& \int p_\theta(x)\log\frac{p_{\theta_1}(x)}{p_{\theta_2}(x)},\\
&=& F(\theta_2)-F(\theta_1)-\inner{\theta_2-\theta_1}{\nabla F(\theta)}.
\end{eqnarray}

In particular, when $\theta=\alpha\theta_1+(1-\alpha)\theta_2$ for $\alpha\in(0,1)$, i.e., $r\propto p_{\theta_1}^\alpha p_{\theta_2}^{1-\alpha}$ 
is a normalized geometric mixture, we have
$$
D_\KL[p_{\theta_1}:p_{\theta_2}|p_{\alpha\theta_1+(1-\alpha)\theta_2}]= F(\theta_2)-F(\theta_1)-\inner{\theta_2-\theta_1}{\nabla F(\alpha\theta_1+(1-\alpha)\theta_2)}.
$$

This $3$-density KL pseudo-divergence
 is related to the Bregman tangent  divergence defined in~\cite{nielsen2018bregman,nielsen2019bregman} but it may be negative.

%%%
\section{Bhattacharyya skewed divergence between nested densities of an exponential family}\label{sec:bhat}
%%%
The Bhattacharyya $\alpha$-skewed divergence~\cite{bhattacharyya1943measure,nielsen2011burbea} between two densities $p(x)$ and $q(x)$ with respect to $\mu$  is defined for a skewing scalar parameter $\alpha\in(0,1)$ as:
$$
D_{\Bhat,\alpha}[p:q]:=-\log \int_\calX p(x)^\alpha q(x)^{1-\alpha}\,\dmu(x),
$$
where $\calX$ denotes the support of the distributions. 
Let $I_\alpha[p:q]:=\int_\calX p(x)^\alpha q(x)^{1-\alpha}\,\dmu(x)$ denote the skewed affinity coefficient so that $D_{\Bhat,\alpha}[p:q]=-\log I_\alpha[p:q]$.
Since $I_\alpha[p:q]=I_{1-\alpha}[q:p]$, we have $D_{\Bhat,\alpha}[p:q]=D_{\Bhat,1-\alpha}[q:p]$.

Consider an exponential family  $\calE=\{p_\theta\}$  with log-normalizer $F(\theta)$.
Then it is well-known that the $\alpha$-skewed Bhattacharyya divergence between two densities of an exponential family amounts to a skewed Jensen divergence~\cite{nielsen2011burbea} (originally called Jensen difference in~\cite{rao1982diversity}):
$$
D_{\Bhat,\alpha}[p_{\theta_1}:p_{\theta_2}]=J_{F,\alpha}(\theta_1:\theta_2),
$$
where the skewed Jensen divergence is defined by
$$
J_{F,\alpha}(\theta_1:\theta_2)=\alpha F(\theta_1)+(1-\alpha) F(\theta_2)-F(\alpha\theta_1+(1-\alpha)\theta_2).
$$
The convexity of the log-normalizer $F(\theta)$ ensures that $J_{F,\alpha}(\theta_1:\theta_2)\geq 0$.
The Jensen divergence can be extended to full real $\alpha$ by rescaling it by $\frac{1}{\alpha(1-\alpha)}$, see~\cite{zhang2004divergence}.

\begin{Remark}
The Bhattacharyya skewed divergence $D_{\Bhat,\alpha}[p:q]$ appears naturally as the negative of the  log-normalizer of the exponential family induced by the exponential arc $\{r_\alpha(x)\: \alpha\in (0,1)\}$ linking two densities $p$ and $q$ with
 $r_\alpha(x)\propto p(x)^{\alpha}q(x)^{1-\alpha}$.
This arc is an exponential family of order $1$:
\begin{eqnarray*}
r_\alpha(x) &=& \exp \left(\alpha\log p(x)+(1-\alpha)\log q(x) -\log Z_\alpha(p:q)\right),\\
&=& \exp \left( \alpha\log \frac{p(x)}{q(x)} -  F_{pq}(\alpha)\right)\, q(x).
\end{eqnarray*}
The sufficient statistic is $t(x)=\frac{p(x)}{q(x)}$, the natural parameter $\alpha\in (0,1)$ and the log-normalizer
$F_{pq}(\alpha)=\log Z_\alpha(p:q)=\log  \int p(x)^{\alpha}q(x)^{1-\alpha}\dmu(x)=-D_{\Bhat,\alpha}[p:q]$.
This shows that $D_{\Bhat,\alpha}[p:q]$ is concave with respect to $\alpha$ since log-normalizers $F_{pq}(\alpha)$ are always convex.
Gr\"unwald called those exponential families: The likelihood ratio exponential families~\cite{grunwald2007minimum}.
\end{Remark}

Now, consider calculating $D_{\Bhat,\alpha}[p_{\theta_1}:q_{\theta_2}]$ where $p_{\theta_1}\in\calE_1$ with $\calE_1$ a nested exponential family of $\calE_2$ and $q_{\theta_2}\in\calE_2=\{q_\theta\}$. 
We have $q_\theta(x)=\frac{Z_1(\theta)}{Z_2(\theta)} p_\theta(x)$, where $Z_1(\theta)$ and $Z_2(\theta)$ are the partition functions of $\calE_1$ and $\calE_2$, respectively.
Thus we have
$$
I_\alpha[p_{\theta_1}:q_{\theta_2}]= \left(\frac{Z_1(\theta_2)}{Z_2(\theta_2)}\right)^{1-\alpha} \, I_\alpha[p_{\theta_1}:p_{\theta_2}],
$$
and the $\alpha$-skewed Bhattacharyya  divergence is
$$
D_{\Bhat,\alpha}[p_{\theta_1}:q_{\theta_2}]=D_{\Bhat,\alpha}[p_{\theta_1}:p_{\theta_2}]-(1-\alpha)(F_1(\theta_2)-F_2(\theta_2)).
$$

Therefore we get
\begin{eqnarray*}
D_{\Bhat,\alpha}[p_{\theta_1}:q_{\theta_2}] &=& J_{F_1,\alpha}(\theta_1:\theta_2)-(1-\alpha)(F_1(\theta_2)-F_2(\theta_2)),\\
&=& \alpha F_1(\theta_1)+(1-\alpha)F_2(\theta_2)-F_1(\alpha\theta_1+(1-\alpha)\theta_2),\\
&=:& J_{F_1,F_2,\alpha}(\theta_1:\theta_2).
\end{eqnarray*}
We call $J_{F_1,F_2,\alpha}(\theta_1:\theta_2)$ the duo Jensen divergence.
Since $F_2(\theta)\geq F_1(\theta)$, we check that
$$
J_{F_1,F_2,\alpha}(\theta_1:\theta_2)\geq J_{F_1,\alpha}(\theta_1:\theta_2)\geq 0.
$$

\begin{figure}%
\centering
\includegraphics[width=0.6\columnwidth]{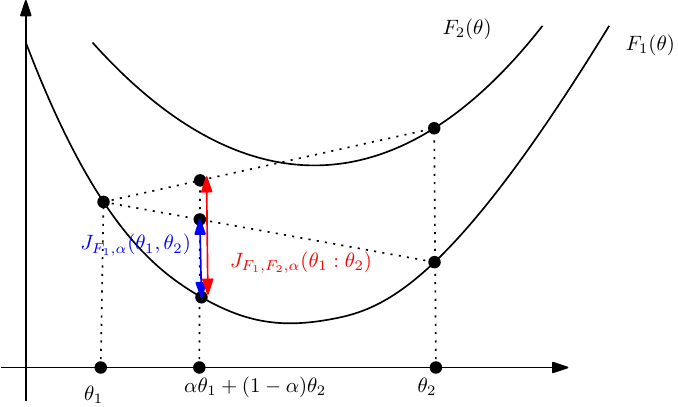}%
\caption{The duo Jensen divergence $J_{F_1,F_2,\alpha}(\theta_1:\theta_2)$ is greater than the Jensen divergence $J_{F_1,\alpha}(\theta_1:\theta_2)$ for $F_2(\theta)\geq F_1(\theta)$.}%
\label{fig:duoJS}%
\end{figure}

Figure~\ref{fig:duoJS} illustrates graphically the duo Jensen divergence $J_{F_1,F_2,\alpha}(\theta_1:\theta_2)$.

\begin{Theorem}\label{thm:BhatNEF}
The $\alpha$-skewed  Bhattacharyya divergence for $\alpha\in(0,1)$ between a truncated density of $\calE_1$ with log-normalizer $F_1(\theta)$ and another density of an exponential family $\calE_2$ with log-normalizer $F_2(\theta)$ amounts to a duo Jensen divergence:
$$
D_{\Bhat,\alpha}[p_{\theta_1}:q_{\theta_2}] = J_{F_1,F_2,\alpha}(\theta_1:\theta_2),
$$
where $J_{F_1,F_2,\alpha}(\theta_1:\theta_2)$ is the duo skewed Jensen divergence induced by two strictly convex functions $F_1(\theta)$ and $F_2(\theta)$ such that $F_2(\theta)\geq F_1(\theta)$:
$$
J_{F_1,F_2,\alpha}(\theta_1:\theta_2)=\alpha F_1(\theta_1)+(1-\alpha)F_2(\theta_2)-F_1(\alpha\theta_1+(1-\alpha)\theta_2).
$$
\end{Theorem}

In~\cite{nielsen2011burbea}, it is reported that
\begin{eqnarray*}
D_\KL[p_{\theta_1}:p_{\theta_2}]&=&B_F(\theta_2:\theta_1),\\
&=&\lim_{\alpha\rightarrow 0} \frac{1}{\alpha} J_{F,\alpha}(\theta_2:\theta_1) = \lim_{\alpha\rightarrow 0} \frac{1}{\alpha} J_{F,1-\alpha}(\theta_1:\theta_2),\\
&=&\lim_{\alpha\rightarrow 0} \frac{1}{\alpha} D_{\Bhat,\alpha}[p_{\theta_2}:p_{\theta_1}]
=\lim_{\alpha\rightarrow 0} \frac{1}{\alpha} D_{\Bhat,1-\alpha}[p_{\theta_1}:p_{\theta_2}].
\end{eqnarray*}

Indeed,  using the first order Taylor expansion of
$$
F(\theta_1+\alpha(\theta_2-\theta_1))=F(\theta_1)+\alpha(\theta_2-\theta_1)^\top\nabla F(\theta_1)
$$ 
when $\alpha\rightarrow 0$,
we check that we have
\begin{eqnarray*}
\frac{1}{\alpha} J_{F,\alpha}(\theta_2:\theta_1) &\approx_{\alpha\rightarrow 0}&
 \frac{F(\theta_1)+\alpha(F(\theta_2)-F(\theta_1))-F(\theta_1+\alpha(\theta_2-\theta_1))}{\alpha},\\
&=& B_F(\theta_2:\theta_1).
\end{eqnarray*}

Moreover, we have
$$
\lim_{\alpha\rightarrow 0} \frac{1}{\alpha} D_{\Bhat,1-\alpha}[p:q] = D_\KL[p:q].
$$

Similarly, we can prove that 
$$
\lim_{\alpha\rightarrow 1} \frac{1}{1-\alpha} J_{F_1,F_2,\alpha}(\theta_1:\theta_2)=B_{F_2,F_1}(\theta_2:\theta_1),
$$
which can be reinterpreted as
$$ 
\lim_{\alpha\rightarrow 1} \frac{1}{1-\alpha} D_{\Bhat,\alpha}[p_{\theta_1}:q_{\theta_2}] = D_\KL[p_{\theta_1}:q_{\theta_2}].
$$

%%%
\section{Sided duo Bregman centroids}
%%%
A centroid of a finite set of $n$ parameters $\{\theta_1,\ldots,\theta_n\}$ with respect to a divergence $D(\cdot:\cdot)$ is defined by the following minimization problem:
\begin{equation}\label{eq:lcentroid}
\min_{\theta} \frac{1}{n}\sum_{i=1}^n D(\theta:\theta_i).
\end{equation}
When $D(\theta:\theta')=\|\theta-\theta'\|^2$m the centroid corresponds to the center of mass $\bar\theta=\frac{1}{n}\sum_{i=1}^n \theta_i$.

Since the divergence $D(\cdot:\cdot)$ may be asymmetric, we may also consider the centroid defined with respect to the dual divergence 
$D^*(\theta:\theta')=D(\theta':\theta)$ (also called the reverse divergence or backward divergence):
\begin{equation}\label{eq:rcentroid}
\min_{\theta} \frac{1}{n}\sum_{i=1}^n D^*(\theta:\theta_i)=\min_{\theta} \frac{1}{n}\sum_{i=1}^n D(\theta_i:\theta).
\end{equation}
We shall call the minimizer(s) of Eq.~\ref{eq:lcentroid} the left-sided centroid(s) and the minimizer(s) of Eq.~\ref{eq:rcentroid} the right-sided centroid(s).

Now, consider the duo Bregman left-sided centroid defined by the following minimization problem:
\begin{equation}\label{eq:dyflcentroid}
\min_{\theta^L} \frac{1}{n}\sum_{i=1}^n B_{F_1,F_2}(\theta^L:\theta_i).
\end{equation}
Using the equivalent duo Fenchel-Young divergence of Eq.~\ref{eq:duobdyf}, the minimization of Eq.~\ref{eq:dyflcentroid} amounts equivalently to
$$
\min_{\theta^L} L(\theta^L):=\sum_{i=1}^n Y_{F_1,F_2^*}(\theta^L:\eta_i)=\sum_{i=1}^n F_1(\theta^L)+F_2^*(\eta_i)-{\theta^L}^\top\eta_i,
$$
where $\eta_i=\nabla F_2(\theta_i)$.
Setting the gradient of $L(\theta^L)$ to zero, we find that $\nabla F_1(\theta^L)=\frac{1}{n}\sum_{i=1}^n \eta_i:=\bar\eta$.
Thus we have
\begin{equation}\label{eq:soldyfcentroidl}
\theta^L=(\nabla F_1)^{-1}\left(\frac{1}{n}\sum_{i=1}^n \nabla F_2(\theta_i)\right).
\end{equation}
When $F_2=F_1=F$ and $F$ is coordinate-wise separable, Eq.~\ref{eq:soldyfcentroidl} is a multivariate quasi-arithmetic mean~\cite{nielsen2011burbea}.

Next, consider the duo Bregman right-sided centroid:
$$
\min_{\theta^R} \sum_{i=1}^n B_{F_1,F_2}(\theta_i:\theta^R).
$$
Using the equivalent duo Fenchel-Young divergence, the minimization amounts equivalently to
$$
\min_{\theta^R} R(\theta^R):= \sum_{i=1}^n Y_{F_1,F_2^*}(\theta_i:\eta^R)= \sum_{i=1}^n F_1(\theta_i)+F_2^*(\eta^R)-{\theta_i}^\top\eta^R,
$$
where $\eta^R=\nabla F_2(\theta^R)$.
Setting the gradient of $R(\theta^R)$ to zero, we find that $\nabla F_2^*(\eta^R)=\frac{1}{n}\sum_{i=1}^n \theta_i$.
That is, we get
$$
\eta^R=(\nabla F_2^*)^{-1}\left(\frac{1}{n}\sum_{i=1}^n \theta_i\right)=\nabla F_2(\theta^R).
$$
Since $(\nabla F_2)^{-1}=(\nabla F_2^*)$, we get
\begin{equation}
\theta^R=\frac{1}{n}\sum_{i=1}^n \theta_i:=\bar\theta.
\end{equation}
Thus the duo Bregman right-sided centroid is always the center of mass.
This generalizes the result of Banerjee et al.~\cite{banerjee2005clustering} (Proposition~1).

Notice that the symmetrized duo Bregman divergence is
\begin{eqnarray*}
S_{F_1,F_2}(\theta_1,\theta_2) &=& B_{F_1,F_2}(\theta_1:\theta_2)+B_{F_1,F_2}(\theta_2:\theta_1),\\
&=& (\theta_1-\theta_2)^\top (\nabla F_2(\theta_1)-\nabla F_2(\theta_2))+ (F_1(\theta_1)-F_2(\theta_1)+F_1(\theta_2)-F_2(\theta_2)).
\end{eqnarray*}
When $F_1=F_2=F$, we recover the ordinary symmetrized Bregman divergence~\cite{nielsen2009sided}:
$$
S_{F}(\theta_1,\theta_2)=S_{F,F}(\theta_1,\theta_2)=
 (\theta_1-\theta_2)^\top (\nabla F(\theta_1)-\nabla F(\theta_2)).
$$

 Clustering with nested exponential family densities (e.g., Zeta and Zipf's laws) has been investigated in~\cite{nielsen2022comparing}.

%%%
\section{Concluding remarks}\label{sec:concl}
%%%

We considered the Kullback-Leibler divergence between two parametric densities $p_\theta\in\calE_1$ and $q_{\theta'}\in\calE_2$ belonging to nested exponential families $\calE_1$ and $\calE_2$, and we showed that their KLD is equivalent to a duo Bregman divergence on swapped parameter order (Theorem~\ref{thm:KLDnestedEF}). 
This result generalizes the study of Azoury and Warmuth~\cite{azoury2001relative}.
The duo Bregman divergence can be rewritten as a duo Fenchel-Young divergence using mixed natural/moment parameterizations of the exponential family densities (Definition~\ref{def:genyf}). This second result generalizes the approach taken in information geometry~\cite{amari1985differential}.
We showed how to calculate the Kullback-Leibler divergence between two truncated normal distributions as a duo Bregman divergence.
More generally, we proved that the skewed Bhattacharyya distance between two parametric nested exponential family densities amount to a duo Jensen divergence (Theorem~\ref{thm:BhatNEF}). 
We show asymptotically that scaled duo Jensen divergences tend to duo Bregman divergences generalizing a result of~\cite{zhang2004divergence,nielsen2011burbea}.
This study of duo divergences induced by pair of generators was motivated by the formula obtained for the  Kullback-Leibler divergence between two densities of two different exponential families originally reported in~\cite{JS-2021} (Eq.~\ref{eq:KLDEFs}).
We called those duo divergences although they are pseudo-divergences since those divergences are always strictly greater than zero when the first generators are strictly majorizing the second generators. 

It is interesting to find applications of the duo Fenchel-Young, Bregman, and Jensen divergences beyond the scope of calculating statistical distances between nested exponential family densities.  
Let us point out that nested exponential families have been seldom considered in the literature (see~\cite{pisano2021occam} for a recent work).
Notice that in~\cite{nielsen2020quasiconvex}, the authors exhibit a relationship between densities with nested supports\footnote{However, those considered parametric densities are not exponential families since their support depend on the parameter.} and  quasi-convex Bregman divergences. 
Recently,  Khan and Swaroop~\cite{emtiyaz2021knowledge} used this duo Fenchel-Young divergence in machine learning for knowledge-adaptation priors in the so-called change regularizer task.

\bibliographystyle{plain}
\bibliography{FYBib.bib}

\end{document}